\documentstyle[aps,twocolumn]{revtex}
\newcommand{\la}{\langle}
\newcommand{\ra}{\rangle}

\input{epsf}

\title{Causal and localizable quantum operations\thanks{CALT-68-2309}}
\author{David Beckman$^{(1)}$\thanks{\tt
beckman@theory.caltech.edu} Daniel Gottesman,$^{(2,3)}$\thanks{\tt gottesma@eecs.berkeley.edu}, M.~A. Nielsen,$^{(4,1)}$\thanks{\tt nielsen@physics.uq.edu.au} and John Preskill$^{(1)}$\thanks{\tt
preskill@theory.caltech.edu}}

\address{$^{(1)}$Institute for Quantum Information, California Institute of Technology, 
Pasadena, CA 91125, USA\\$^{(2)}$Microsoft Corporation, One Microsoft Way, Redmond, WA 98052, USA\\$^{(3)}$Computer Science Division, EECS, University of California, Berkeley, CA 94720, USA\\$^{(4)}$Center for Quantum Computer Technology, University of Queensland 4072, Australia}

\begin{document}

\maketitle

\begin{abstract}
We examine constraints on quantum operations imposed by relativistic causality.  A bipartite superoperator is said to be {\em localizable} if it can be implemented by two parties (Alice and Bob) who share entanglement but do not communicate; it is {\em causal} if the superoperator does not convey information from Alice to Bob or from Bob to Alice. We characterize the general structure of causal complete measurement superoperators, and exhibit examples that are causal but not localizable. We construct another class of causal bipartite superoperators that are not localizable by invoking bounds on the strength of correlations among the parts of a quantum system. A bipartite superoperator is said to be {\em semilocalizable} if it can be implemented with one-way quantum communication from Alice to Bob, and it is {\em semicausal} if it conveys no information from Bob to Alice. We show that all semicausal complete measurement superoperators are semilocalizable, and we establish a general criterion for semicausality. In the multipartite case, we observe that a measurement superoperator that projects onto the eigenspaces of a stabilizer code is localizable.
\end{abstract}

\parskip=5pt
\section{Introduction}

What are the {\em observables} of a relativistic quantum theory? Standard wisdom holds that any self-adjoint operator that can be defined on a spacelike slice through spacetime is measurable in principle. But in fact, measuring a typical such operator is forbidden by relativistic causality, and hence impossible.

More generally, it is often stated that the possible ways that the state of a quantum system can change are described by  {\em quantum operations} -- completely positive trace non-increasing linear maps of density operators to density operators \cite{nielsen_chuang,preskill229}. But in a relativistic quantum theory, typical operations would allow superluminal signaling, and are therefore unphysical. 

Relativistic quantum field theory allows us to identify an algebra of observables that is compatible with the causal structure of spacetime \cite{haag}. Despite this marvelous achievement, puzzles and open questions remain. Our objective in this paper is to understand better the restrictions on operations that are imposed by special relativity. Mostly, we will consider a simplified version of the problem in which the physical system is divided into two separated parts: part $A$, which is controlled by a party that we will call Alice, and part $B$, which is controlled by Bob. Initially, Alice and Bob share a joint quantum state whose density operator $\rho_{AB}$ is not known, and they wish to transform the state to ${\cal E}(\rho_{AB})$, where ${\cal E}$ is a specified operation.

If Alice and Bob were able to communicate by sending quantum information back and forth, then they would be able to apply {\em any} operation ${\cal E}_{AB}$ to their state. We want to determine what operations they can implement if {\em no} communication (quantum or classical) is permitted. In a relativistic setting, these are the operations that can be realized if Alice's action and Bob's action are spacelike-separated events. We will, though, permit Alice and Bob to make use of a shared entangled {\em ancilla} state that might have been prepared earlier and distributed to them. 

While Alice and Bob are permitted to perform measurements, Bob cannot know the outcome of Alice's measurement, and Alice cannot know the outcome of Bob's. Therefore, we will largely restrict our attention to {\em trace-preserving} quantum operations, also known as {\em superoperators}, where no postselection of the quantum state based on the measurement outcome is allowed. We say that a bipartite superoperator is {\em localizable} if it can be implemented by Alice and Bob acting locally on the shared state and the shared ancilla, without any communication from Alice to Bob or Bob to Alice. 

Another important concept is that of a {\em causal operation}. We say that an operation is causal if it does not allow either party to send a signal to the other. More precisely, imagine that Bob applies a local superoperator ${\cal B}$ to his half of the state he shares with Alice just before the global operation ${\cal E}$ acts on the joint system, and that Alice makes a local measurement on her half just after ${\cal E}$ acts. If Alice's measurement can acquire any information about what operation was applied by Bob, then we say that Bob can signal Alice. The operation is causal if no such signaling is possible in either direction. 

Entanglement shared by Alice and Bob cannot be used to send a superluminal signal from Alice to Bob or from Bob to Alice. Therefore, any localizable superoperator is surely a causal superoperator. What about the converse? It might seem reasonable to expect that {\em any} operation, if it respects the principle that information cannot propagate outside the forward light cone, should be physically realizable in principle. However, we will show otherwise by exhibiting some examples of superoperators that are causal but not localizable.

We obtain weaker notions of localizability and causality by considering communication in just one direction. We say that a superoperator is {\em semilocalizable} if it is possible to implement it with {\em one-way} quantum communication from Alice to Bob. Such operations are physically realizable if Bob's action takes place in the future light cone of Alice's action. Similarly, we say that an operation is {\em semicausal} if it does not allow Bob to send a signal to Alice. Obviously, a semilocalizable superoperator is semicausal -- communication from Alice to Bob cannot help Bob to send a message to Alice.  What about the converse? If one believes that causality is a very special property of operations that is not likely to hold ``by accident,'' then it is natural to formulate the following conjecture, suggested to us by DiVincenzo \cite{divincenzo}:

\noindent {\bf Conjecture} {\em Every semicausal superoperator is semilocalizable.}

\noindent We will prove this conjecture for the special case of complete orthogonal measurement superoperators. Whether it holds in general remains an open question.

The problem of characterizing what measurements are possible was raised by Dirac \cite{dirac}, and interesting examples of impossible measurements were pointed out in \cite{wigner,araki,nielsen_imp}. That relativistic causality may restrict the operators that can be measured in a field theory was first emphasized by Landau and Peierls \cite{landau} (though their particular concerns were well answered by Bohr and Rosenfeld \cite{bohr}). More recently, these restrictions have been noted by a variety of authors \cite{ahar86,ahar80,ahar81,sorkin,popescu,beckman}. In \cite{preskill}, we have addressed some particular causality issues that arise in non-Abelian gauge theories.

To apply our notion of localizability to quantum field theory, we must adopt the convenient fiction that the entangled ancilla is an external probe not itself described by the field theory, and that its local coupling to the fields is completely adjustable. This idealization is highly questionable in a quantum theory of gravity, and even for quantum field theory on flat spacetime it is open to criticism. In particular, field variables in spatially adjacent regions are inevitably entangled \cite{haag}, so that no strict separation between field and ancilla variables is really possible. On the other hand, if the probe variables are ``heavy'' with rapidly decaying correlations and the field variables are ``light,'' then our idealization is credible and worthy of study.

Should the conjecture that semicausality implies semilocalizability prove to be true, then we will have a general and powerful criterion for deciding if a superoperator can be executed with one-way communication.  Even so, we will lack a fully satisfactory way of characterizing the observables of a relativistic quantum theory, as no communication is possible if an operation is carried out on a spacelike slice. The existence of causal quantum operations that are not localizable establishes a perplexing gap between what is {\em causal} and what is {\em local}.

In \S II, we formulate precise definitions of causal, semicausal, localizable, and semilocalizable, and we point out a large class of localizable superoperators characterized by local stabilizer groups. We describe the general structure of semicausal and causal complete measurement superoperators in \S III, show that semicausal complete measurement superoperators are semilocalizable in \S IV, and exhibit some causal complete measurement superoperators that are not localizable in \S V. In \S VI, we exploit bounds on the strength of quantum correlations to construct another class of causal superoperators that are not localizable, and we note a connection between localizability and quantum communication complexity. We prove in \S VII that a semicausal unitary transformation must be a tensor product.  Some further criteria for semicausality are developed in \S VIII, and \S IX contains some concluding comments. Proofs of two of our theorems are included as appendices.

\section{Causality and Localizability}
In this section, we formally define the properties of quantum operations that we wish to explore -- causality, semicausality, localizability, and semilocalizability -- and we discuss some examples that illustrate these concepts. 
\subsection{Causality}
Any permissible way in which the state of a quantum system can change is described by a quantum operation, a completely positive trace-nonincreasing linear map of density operators to density operators.  An important special case is a trace-preserving map, or {\em superoperator}. A superoperator ${\cal E}$ can be interpreted as a generalized measurement with an unknown outcome; its action on a density operator $\rho$ has an operator-sum representation
\begin{equation}
\label{operator_sum}
{\cal E}(\rho)=\sum_\mu M_\mu \rho M_\mu^\dagger~,
\end{equation}
where the operation elements $M_\mu$ obey the normalization condition
\begin{equation}
\label{operator_normalization}
\sum_\mu M_\mu^\dagger M_\mu= I~.
\end{equation}
An operation is a generalized measurement in which a particular outcome has been selected, but the density operator has not been renormalized. It too can be represented as in eq.~(\ref{operator_sum}), but where the sum over $\mu$ is restricted to a subset of a set of operators obeying eq.~(\ref{operator_normalization}) -- that is, the eigenvalues of $\sum_\mu M_\mu^\dagger M_\mu$ are no greater than 1. For a general operation, ${\rm tr}~{\cal E}(\rho)$ can be interpreted as the probability of the observed outcome.

Every superoperator has a unitary representation. To implement the superoperator ${\cal E}_S$ acting on Hilbert space ${\cal H}_S$, we can introduce an ancilla with Hilbert space ${\cal H}_R$, prepare a pure state $|\psi\rangle\in {\cal H}_R$ of the ancilla, perform a unitary transformation $U$ on ${\cal H}_S\otimes{\cal H}_R$, and then discard the ancilla:
\begin{equation}
{\cal E}_S(\rho_S) = {\rm tr}_R~\big[ U\left(\rho_S\otimes |\psi\rangle_R {}_R\langle \psi|\right) U^\dagger\big]~.
\end{equation}
A general operation has a similar representation, except that after $U$ is applied, a (not necessarily complete) orthogonal measurement is carried out on the ancilla and a particular result is selected (without renormalizing the density operator).

\begin{figure}
\begin{center}
\leavevmode
\epsfxsize=3in
\epsfbox{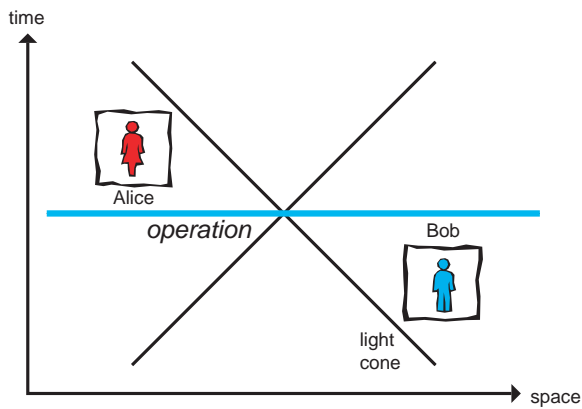}
\end{center}
\caption{An operation on a time slice. If the operation allows spacelike-separated Alice and Bob to communicate, then it is not {\em causal} and hence not physically implementable.}
\label{fig:lightcone}
\end{figure}

But what quantum operations are physically possible? The general answer is not known, but it {\em is} known that many operations of the form eq.~(\ref{operator_sum}) are unphysical because they run afoul of relativistic causality \cite{ahar86,ahar80,ahar81,sorkin,popescu,beckman}. Consider, as in Fig.~\ref{fig:lightcone}, two parties Alice and Bob who perform spacelike-separated actions. Just prior to the implementation of the operation ${\cal E}$, Bob performs a local operation on the degrees of freedom in his vicinity, and just after the implementation of ${\cal E}$, Alice performs a local measurement of the degrees of freedom in her vicinity. If Alice is able to acquire any information about what local operation Bob chose to apply, then Bob has successfully sent a superluminal signal to Alice. 

If a bipartite operation ${\cal E}$ does not enable such superluminal signaling from Bob to Alice, then we will say that ${\cal E}$ is {\em semicausal}. If ${\cal E}$ does not allow signaling in either direction, we will say that ${\cal E}$ is {\em causal}. (In our discussions of semicausality, we will normally adopt the convention that Bob is the party attempting to send the signal and Alice is the party attempting to receive it. This somewhat perverse convention is chosen in order to be consistent with the definition of semilocalizable that is introduced below.) An operation that is not semicausal is said to be {\em acausal}. A causal operation will sometimes be called {\em fully causal}, when we wish to emphasize the contrast with semicausality.

Alice's part of the bipartite system will be called ``system $A$'', with Hilbert space ${\cal H}_A$, and Bob's half is ``system $B$'' with Hilbert space ${\cal H}_B$. We consider a quantum operation ${\cal E}$ that acts on states in ${\cal H}_A\otimes {\cal H}_B$. Alice might have under her control not only $A$, but also an ancilla system $R$ with Hilbert space ${\cal H}_R$, and Bob might control an ancilla system $S$ with Hilbert space ${\cal H}_S$. Suppose that Bob wants to attempt to exploit the operation ${\cal E}$ to send a signal to Alice. Alice and Bob could share an initial density operator $\rho_{RABS}$ living in ${\cal H}_R\otimes {\cal H}_A\otimes {\cal H}_B\otimes {\cal H}_S$, and Bob could apply a superoperator ${\cal B}_{BS}$ to his half of the state. (Bob is restricted to a superoperator, rather than a trace-decreasing operation, because Alice is unaware of the outcome of any measurement performed by Bob.) Then after the operation ${\cal E}$ is applied, we obtain Alice's final density operator by tracing over Bob's system and ancilla
\begin{equation}
\rho_{RA}= {\rm tr}_{BS}\big[{\cal E}_{AB}\big( (I_{RA}\otimes {\cal B}_{BS}(\rho_{RABS}) \big)\big]~.
\end{equation}
Here by ${\cal E}_{AB}$ we mean the operation $I_R\otimes {\cal E}_{AB}\otimes I_S$ that acts trivially on the ancillas. Note that if ${\cal E}$ is not trace preserving, then $\rho_{RA}$ might not be normalized, in which case Alice's final state is the normalized density operator $\rho_{RA}/{\rm tr} ~\rho_{RA}$. Finally, Alice performs a measurement of this state.

If Alice's state depends at all on the superoperator ${\cal B}$ that Bob applies, then the mutual information of Bob's superoperator and Alice's measurement is nonzero. Hence Bob can transmit classical information to Alice over a noisy channel with nonzero capacity; that is, Bob can signal Alice. We arrive then, at this

\noindent {\bf Definition} {\em A bipartite operation ${\cal E}$ is {\bf semicausal} if and only if $(\rho_{RA}/{\rm tr}~\rho_{RA})$ is independent of Bob's superoperator ${\cal B}$ for all possible choices of the shared initial state $\rho_{RABS}$.} 

\noindent (Excluded from consideration is the case $\rho_{RA}=0$, corresponding to an outcome that occurs with probability zero.)

This criterion for semicausality is rather unwieldy; fortunately it can be simplified. One useful observation is that, while in our definition of semicausality we allowed the initial state $\rho_{RABS}$ shared by Alice and Bob to be entangled, we could without loss of generality restrict their initial state to be a product state.

Suppose that a {\em superoperator} ${\cal E}$ is not semicausal. Then there is an initial state $\rho_{RABS}$ shared by Alice and Bob, and a superoperator ${\cal B}$ that can be applied by Bob, such that
\begin{eqnarray}
\label{not_one_causal}
&&{\rm tr}_{BS}\big[{\cal E}_{AB}(\rho_{RABS})\big]\nonumber\\
&&\ne{{\rm tr}_{BS}\big[{\cal E}_{AB}\big(I_{RA}\otimes {\cal B}_{BS}(\rho_{RABS})\big)\big]
}~.
\end{eqnarray}
Now any bipartite density operator can be expanded as
\begin{eqnarray}
  \rho_{RABS} = \sum_\mu \lambda_\mu \rho_\mu \otimes \sigma_\mu~,
\end{eqnarray}
where $\rho_\mu$ and $\sigma_\mu$ are density operators of Alice's and Bob's
systems (including ancillas), respectively, and the $\lambda_\mu$'s  are nonvanishing real numbers. Of course, if the $\lambda_\mu$'s were all positive, then $\rho_{RABS}$ would be a separable state.  But if we allow the $\lambda_\mu$'s to be negative, then such an expansion exists for any state.

Since the superoperator ${\cal E}$ is linear, we may rewrite eq.~(\ref{not_one_causal}) as
\begin{eqnarray}
&&\sum_\mu \lambda_\mu~ {\rm tr}_{BS}\big[{\cal E}\big(\rho_\mu \otimes \sigma_\mu\big)\big]\nonumber\\
&&\ne\sum_\mu \lambda_\mu ~ {{\rm tr}_{BS}\big[{\cal E}\big(\rho_\mu \otimes {\cal B}(\sigma_\mu)\big)\big]} ~,
\end{eqnarray}
which can be satisfied only if 
\begin{eqnarray}
 {\rm tr}_{BS}\big[{\cal E}\big(\rho_\mu \otimes \sigma_\mu\big)\big]\ne
 {{\rm tr}_{BS}\big[{\cal E}\big(\rho_\mu \otimes {\cal B}(\sigma_\mu)\big)\big]}~,
\end{eqnarray}
for at least one $\mu$. Therefore, if Alice and Bob prepare the appropriate product state $\rho_\mu \otimes \sigma_\mu$, ${\cal E}$ allows Bob to signal Alice. Furthermore, since each of $\rho_\mu$ and $\sigma_\mu$ and ${\cal B}(\sigma_\mu)$ can be realized as an ensemble of pure states, there is a signaling protocol in which Alice's and Bob's initial states are pure.

Once we recognize that there is a signaling protocol such that the initial state is a product state, we can see that the ancillas are superfluous. Bob sends his signal by choosing one of the two pure states $|\psi\rangle_{BS}$, $|\psi'\rangle_{BS}$; since tracing over $S$ commutes with ${\cal E}$, we can just as well say that Bob starts with a mixed state $\rho_B$ or $\rho'_B$ of system $B$ alone. Furthermore, if Bob can signal Alice by preparing one of $\rho_B$, $\rho'_B$, then he must be able to do it by preparing pure states that arise in the ensemble realizations of these density operators.

Finally, if signaling is possible, then Alice can receive the signal by preparing an initial pure state $|\varphi\rangle_{RA}$. Bob's action, together with ${\cal E}$, subjects system $A$ to one of two possible operations, resulting in two distinguishable final states. But for these final states of $RA$ to be distinguishable, the two operations must produce different outcomes acting on at least one of the {\em pure} states of $A$ appearing in the Schmidt decomposition of $|\varphi\rangle_{RA}$. Therefore, Alice could just as well dispense with $R$, and prepare an initial pure state $|\varphi\rangle_A$ of $A$ alone. 

Thus we have proved

\noindent{\bf Theorem 1} {\em If the bipartite superoperator ${\cal E}$ is not semicausal, then signaling is possible with pure initial states and without ancillas: there are pure states $|\psi\rangle_B, |\psi'\rangle_B\in {\cal H}_B$ and $|\varphi\rangle_A\in {\cal H}_A$ such that
\begin{eqnarray}
&&{\rm tr}_{B}\big[{\cal E}\big( (|\varphi\rangle\langle\varphi|)_A~\otimes ~(|\psi\rangle\langle \psi|)_B\big)\big]\nonumber\\
&&\ne{\rm tr}_{B}\big[{\cal E}\big( (|\varphi\rangle\langle\varphi|)_A~\otimes ~(|\psi'\rangle\langle \psi'|)_B\big)\big]~.
\end{eqnarray}
}

We note that semicausal superoperators form a convex set. If each ${\cal E}_a$ is a superoperator, then so is the combination
\begin{equation}
{\cal E}=\sum_a p_a {\cal E}_a ~,
\end{equation}
where the $p_a$'s are nonnegative and sum to 1. It follows from the linearity of the ${\cal E}_a$'s and the definition of semicausality that ${\cal E}$ is semicausal if each ${\cal E}_a$ is. 

A somewhat less obvious property is that the semicausal superoperators form a semigroup -- a composition of semicausal operations is semicausal. This follows from

\noindent{\bf Theorem 2} {\em Suppose that ${\cal E}$ is a semicausal bipartite superoperator, and that the two bipartite density operators $\rho$ and $\sigma$ satisfy ${\rm tr}_B\rho={\rm tr}_B\sigma$. Then ${\rm tr}_B{\cal E}(\rho)={\rm tr}_B {\cal E}(\sigma)$.}

\noindent {\bf Proof}: First we note that it is possible to choose a basis of linearly independent operators acting on a Hilbert space ${\cal H}$, such that each element of the basis is a one-dimensional projector. Let $\{P_\mu\}$ denote such a basis for Alice's Hilbert space ${\cal H}_A$ and let $\{Q_\mu\}$ denote such a basis for Bob's Hilbert space ${\cal H}_B$. Then we may expand the bipartite density operators $\rho$ and $\sigma$ as
\begin{eqnarray}
\rho &=& \sum_{\mu\nu}\alpha_{\mu\nu}P_\mu\otimes Q_\nu~,\nonumber\\
\sigma &=& \sum_{\mu\nu}\alpha'_{\mu\nu}P_\mu\otimes Q_\nu~.
\end{eqnarray}
Since ${\rm tr}~Q_\nu=1$, the property ${\rm tr}_B\rho={\rm tr}_B\sigma$ can be rewritten as
$\sum_{\mu\nu}\alpha_{\mu\nu}P_\mu= \sum_{\mu\nu}\alpha'_{\mu\nu}P_\mu$, which implies, since the $P_\mu$'s are a basis, that
\begin{equation}
\label{alpha_sum}
\sum_\nu\alpha_{\mu\nu}=\sum_\nu\alpha'_{\mu\nu}~,
\end{equation}
for each $\mu$.

Applying the superoperator ${\cal E}$, we find that
\begin{eqnarray}
{\cal E}(\rho) &=& \sum_{\mu\nu}\alpha_{\mu\nu}{\cal E}(P_\mu\otimes Q_\nu)~,\nonumber\\
{\cal E}(\sigma) &=& \sum_{\mu\nu}\alpha'_{\mu\nu}{\cal E}(P_\mu\otimes Q_\nu)~.
\end{eqnarray}
Furthermore, since each $Q_{\nu}$ is a pure state, a unitary transformation applied by Bob can transform any one of the $Q_{\nu}$'s to any other; therefore the semicausality of ${\cal E}$ implies that the operator ${\rm tr}_B{\cal E}(P_\mu\otimes Q_\nu)$ is independent of $\nu$. Denoting this operator by $R_\mu$, we have 
\begin{eqnarray}
{\rm tr}_B{\cal E}(\rho) &=& \sum_{\mu\nu}\alpha_{\mu\nu}R_\mu~,\nonumber\\
{\rm tr}_B{\cal E}(\sigma) &=& \sum_{\mu\nu}\alpha'_{\mu\nu}R_\mu~;
\end{eqnarray}
eq.~(\ref{alpha_sum}) then implies that ${\rm tr}_B{\cal E}(\rho)={\rm tr}_B{\cal E}(\sigma)$. This completes the proof of Theorem 2.

The semigroup property of semicausal superoperators is a simple 

\noindent{\bf Corollary} {\em If ${\cal E}_1$ and ${\cal E}_2$ are semicausal bipartite superoperators, then their composition ${\cal E}_2\circ {\cal E}_1$ is also semicausal.}

\noindent{\bf Proof}: Suppose that the bipartite density operator $\rho$ can be transformed to $\sigma$ by a superoperator applied by Bob. Then the semicausality of ${\cal E}_1$ implies that 
${\rm tr}_B{\cal E}_1(\rho)={\rm tr}_B{\cal E}_1(\sigma)$, and Theorem 2 applied to ${\cal E}_2$ implies that 
${\rm tr}_B {\cal E}_2({\cal E}_1(\rho))={\rm tr}_B {\cal E}_2({\cal E}_1(\sigma))$; therefore ${\cal E}_2\circ{\cal E}_1$ is semicausal.

\subsection{Localizability}

Physics is local. If a physical system has many parts that are remote from one another, then the evolution of the system is governed by local ``parties'' that act on the different parts of the system separately. In particular, since communication outside the light cone is impossible, the operations that can be applied to a physical system at a fixed time are those that require no communication among the local parties. We call an operation of this type {\em localizable}.

Although they are not permitted to communicate, the parties are free to exploit any resources that might have been prepared in advance. In particular, they are permitted to have a shared ancilla that might be in an entangled quantum state, and to consume their shared entanglement in the course of executing their operation.

In the case of a system with two parts, one controlled by Alice, the other by Bob, these considerations motivate this

\noindent {\bf Definition} {\em A bipartite superoperator ${\cal E}$ is {\bf localizable} if and only if 
\begin{eqnarray}
{\cal E}(\rho_{AB})= {\rm tr}_{RS} \big[{\cal A}_{RA}\otimes {\cal B}_{BS}(\rho_{AB}\otimes \rho_{RS})\big]
\end{eqnarray}
for some shared ancilla state $\rho_{RS}$ and local superoperators ${\cal A}_{RA}$, ${\cal B}_{BS}$.}

\noindent In fact, by extending the ancilla, the state $\rho_{RS}$ can be ``purified'' and the local superoperators can be replaced by unitary transformations; thus without loss of generality we may use instead the 

\noindent {\bf Definition} {\em A bipartite superoperator ${\cal E}$ is {\bf localizable} if and only if 
\begin{eqnarray}
&&{\cal E}(\rho_{AB})\nonumber\\
&&\quad = {\rm tr}_{RS} \big[U_{RA}\otimes V_{BS}\left(\rho_{AB}\otimes \rho_{RS}\right)U^\dagger_{RA}\otimes V^\dagger_{BS}\big]
\end{eqnarray}
for some shared ancilla pure state $\rho_{RS}$ and local unitary transformations $U_{RA}$ and $V_{BS}$.}

Localizable superoperators form a convex set. To see this, we note that with shared entanglement, Alice and Bob can simulate shared randomness (a weaker resource). For example, suppose they share an ancilla prepared in the state
\begin{equation}
|\Phi\rangle_{RS}=\sum_a\sqrt{p_a} ~ |a\rangle_R\otimes |a\rangle_S~,
\end{equation}
where $\{|a\rangle_R\}$ is an orthonormal basis for Alice's Hilbert space ${\cal H}_R$, $\{|a\rangle_S\}$ is an orthonormal basis for Bob's Hilbert space ${\cal H}_S$, and the $p_a$'s are nonnegative real numbers that sum to one. Then if Alice and Bob both perform measurements that project onto these bases, each obtains the outcome $|a\rangle$ with probability $p_a$. Now let $\{{\cal E}_a\}$ be a set of localizable operations. Alice and Bob can consult their shared randomness, and then carry out a local protocol that applies the operation ${\cal E}_a$ with probability $p_a$, thus achieving a local implementation of the convex sum $\sum_a p_a{\cal E}_a$.

Of course, a superoperator is surely localizable if it is a tensor product of superoperators applied by Alice and by Bob, ${\cal E}={\cal E}_A\otimes {\cal E}_B$. By convexity, any  superoperator of the form 
\begin{equation}
{\cal E}=\sum_a p_a {\cal E}_{A,a}\otimes {\cal E}_{B,a}
\end{equation}
is also localizable. There are some less obvious examples of localizable operations, as we will soon see.

We are also interested in bipartite operations that can be implemented with communication in just one direction. We call such operations {\em semilocalizable}. In our discussions of semilocalizability, we will normally adopt the convention that Alice is permitted to send quantum information to Bob, but Bob cannot send anything to Alice. A semilocalizable operation is one that can be performed in principle if Bob's action is in the forward light cone of Alice's action.

We could, equivalently, provide Alice and Bob with prior shared entanglement, and restrict them to classical communication. These two notions of semilocalizability are equivalent because prior entanglement and classical communication from Alice to Bob enable Alice to teleport quantum information to Bob. Conversely, if Alice can send qubits to Bob, she can establish shared entanglement with him, and she can send him classical messages.

An operation that is not semilocalizable is said to be {\em unlocalizable}. A localizable operation will sometimes be called {\em fully localizable}, when we wish to emphasize the contrast with semilocalizability.

If Alice can send quantum information to Bob, then Alice and Bob both have access to the same ancilla: Alice performs a local operation on the ancilla and her half of the shared state, she sends the ancilla to Bob, and then Bob performs a local operation on the ancilla and his half of the state. Thus we arrive at this

\noindent {\bf Definition} {\em A bipartite operation ${\cal E}$ is {\bf semilocalizable} if and only if 
\begin{eqnarray}
{\cal E}(\rho_{AB})= {\rm tr}_{R} \big[ \big({\cal B}_{BR}\circ{\cal A}_{RA}\big)(\rho_{AB}\otimes \rho_{R})\big]
\end{eqnarray}
for some ancilla state $\rho_{R}$, where ${\cal A}_{RA}$ is an operation and ${\cal B}_{BR}$ is a superoperator.}

\noindent Note that the product ${\cal B}_{BR}\circ{\cal A}_{RA}$ is a composition of operations (with Alice's operation acting first), not a tensor product; the operations do not commute because they act on the same ancilla. We have allowed Alice to apply an operation that is not necessarily trace preserving, since Alice can perform a measurement whose outcome is known to both Alice and Bob, but Bob is restricted to a superoperator because the outcome of a measurement that he performs is not known by Alice. If the operation ${\cal E}$ is a superoperator, then so must be ${\cal A}_{RA}$, and in fact we can take ${\cal B}_{BR}$ and ${\cal A}_{RA}$ to be unitary transformations without loss of generality. 

An obvious consequence of this definition is that semilocalizable (or localizable) superoperators form a semigroup: ${\cal E}_2\circ {\cal E}_1$ is semilocalizable if ${\cal E}_1$ and ${\cal E}_2$ are both semilocalizable. 

\subsection{Orthogonal measurement superoperators}
One of our goals is to characterize the {\em observables} of a relativistic quantum theory: what self-adjoint operators are really measurable?

When we speak of a ``measurement'' of an observable whose support is on a time slice, we need not require that the measurement outcome be instantaneously known by anyone. We might imagine instead that many parties distributed over the slice perform simultaneous local operations. Later the data collected by the parties can be assembled and processed at a central location to determine the measurement result.

Then we may say that the operation performed on the slice is a measurement with an unknown outcome. If $\{E_a\}$ is the set of orthogonal projectors onto the eigenspaces of the observable, the effect of this operation on a density operator $\rho$ is
\begin{equation}
\label{meas_so}
\rho\to {\cal E}(\rho)=\sum_a E_a\rho E_a~.
\end{equation}
We will call a (trace-preserving) operation of this form an {\em orthogonal measurement superoperator}. In the special case where each projector $E_a$ is one-dimensional, it is a {\em complete orthogonal measurement superoperator}, or just a complete measurement superoperator. The causality and localizability properties of complete measurement superoperators will be extensively discussed in the next few sections.

First, let's clarify the concept of semicausality by pointing out an example, noted by Sorkin \cite{sorkin}, of an incomplete measurement superoperator that is not semicausal. It is a two-outcome incomplete Bell measurement performed on a pair of qubits. The orthogonal projectors corresponding to the two outcomes are
\begin{eqnarray}
\label{sorkin}
E_1&=&|\phi^+\rangle\langle \phi^+|~,\nonumber\\
E_2&=&I-|\phi^+\rangle\langle \phi^+|~,
\end{eqnarray}
where $|\phi^+\rangle= (|00\rangle +|11\rangle)/\sqrt{2}$. 
Suppose that the initial pure state shared by Alice and Bob is $|01\rangle_{AB}$. This state is orthogonal to $|\phi^+\rangle$, so that outcome 2 occurs with probability one, and the state is unmodified by the superoperator. Afterwards Alice still has a density operator $\rho_A=|0\rangle\langle 0|$.

But what if, before the superoperator acts, Bob performs a unitary that rotates the state to $|00\rangle_{AB}$? Since this state is an equally weighted superposition of $|\phi^+\rangle$ and $|\phi^-\rangle= (|00\rangle - |11\rangle)/\sqrt{2}$, the two outcomes occur equiprobably, and in either case the final state is maximally entangled, so that Alice's density operator afterwards is $\rho_A=I/2$, where $I$ denotes the identity. Alice can make a measurement that has a good chance of distinguishing the density operators $|0\rangle\langle 0|$ and $I/2$, so that she can decipher a message sent by Bob. By a similar method, Alice can send a signal to Bob. The measurement superoperator is acausal.

On the other hand, some orthogonal measurement superoperators are causal.
For example, measurement of a tensor product observable $A\otimes B$ is obviously causal. -- Alice and Bob can induce decoherence in the basis of eigenstates of a tensor product through only local actions. But there are other examples of causal measurement superoperators that are a bit less obvious. One is complete Bell measurement, {\it i.e.} decoherence in the Bell basis 
\begin{eqnarray}
|\phi^{\pm}\rangle = {1\over\sqrt{2}}\left(|00\rangle\pm |11\rangle\right)~,\nonumber\\
|\psi^\pm\rangle= {1\over\sqrt{2}}\left(|01\rangle\pm |10\rangle\right)~.
\end{eqnarray}
No matter what Bob does, the shared state after Bell measurement is maximally entangled, so that Alice always has $\rho_B=I/2$, and she can't extract any information about Bob's activities. 

Though Bell measurement is a causal operation, it is not something that Alice and Bob can achieve locally without additional resources. But the Bell measurement superoperator is localizable -- Alice and Bob can implement it if they share an entangled ancilla. In fact, shared randomness, a weaker resource than entanglement, is sufficient for this purpose \cite{bdsw}. Suppose that Alice and Bob share a pair of qubits, and also share a string of random bits. At a particular time, they both consult two bits of the  random string; depending on whether they read $00,01,10,$ or $11$, they both apply the unitary operator $I,X,Z,$ or $Y$, where $I$ is the identity, and $\{X,Y,Z\}$ are the $2\times 2$ Pauli matrices
\begin{equation}
X=\pmatrix{0&1\cr 1&0}~,\quad Y=\pmatrix{0&-i\cr i&0}~,\quad Z=\pmatrix{1&0\cr 0&-1}~.
\end{equation}
Together, then, Alice and Bob apply the superoperator
\begin{eqnarray}
{\cal E}(\rho)= && {1\over 4}[(I\otimes I)\rho (I\otimes I) + (X\otimes X)\rho (X\otimes X) \nonumber\\
&+& (Y\otimes Y)\rho (Y\otimes Y) + (Z\otimes Z)\rho (Z\otimes Z)]~.
\end{eqnarray}
The four Bell states are simultaneous eigenstates of $X\otimes X$ and $Z\otimes Z$ (and therefore also $Y\otimes Y= - (X\otimes X)\cdot (Z\otimes Z)$) with eigenvalues $\pm 1$: $Z\otimes Z$ specifies a parity bit that distinguishes $\phi$ from $\psi$ and $X\otimes X$ specifies a phase bit that distinguishes $+$ from $-$. Hence we easily verify that ${\cal E}$ preserves each of the four Bell basis states, and annihilates all the terms in $\rho$ that are off the diagonal in the Bell basis. 

The Bell measurement superoperator can be viewed as a refinement, or ``completion,'' of the 
acausal incomplete measurement superoperator of eq.~(\ref{sorkin}) -- that is, Bell measurement is obtained by resolving the three-dimensional projector $E_2$ of eq.~(\ref{sorkin}) into a sum of three mutually orthogonal one-dimensional projectors.  Thus, this example illustrates that a completion of an acausal measurement superoperator can sometimes be causal. On the other hand, there are other ways of refining the superoperator of eq.~(\ref{sorkin}) that yield acausal complete measurement superoperators. For example, the two-qubit superoperator with projectors
\begin{eqnarray}
E_1 &=&|\phi^+\rangle\langle \phi^+|~,\nonumber\\
E_2&=&|\phi^-\rangle\langle \phi^-|~,\nonumber\\
E_3&=&|01\rangle\langle 01|~,\nonumber\\
E_4&=&|10\rangle\langle 10|~,
\end{eqnarray}
is easily seen to be acausal by applying the criterion of Theorem 3 below. In fact, it is a general feature that if an orthogonal measurement superoperator ${\cal E}$ allows Bob to signal Alice, then there exists a completion of ${\cal E}$ that also allows signaling, with the same signal states $|\psi\rangle_B$ and $|\psi'\rangle_B$. This result is proved in \cite{beckman}.

Since the Bell measurement superoperator can be implemented with shared randomness, one may wonder whether shared randomness is sufficient for the implementation of arbitrary localizable superoperators. But it is easy to think of localizable superoperators for which shared randomness does not suffice -- shared entanglement is necessary. For example, Alice and Bob can locally perform a two-qubit operation in which they throw their qubits away, and replace them with a  $|\phi^+\rangle$ drawn from their shared ancilla. This operation can turn a product state into an entangled state, which would be impossible with local operations and shared randomness alone. We will discuss another example of a localizable superoperator that cannot be implemented with shared randomness in \S VI.

\begin{figure}[h]
\centering
\begin{picture}(230,106)

\put(44,0){\makebox(10,12){($a$)}}
\put(184,0){\makebox(10,12){($b$)}}

\put(0,34){\makebox(10,12){$|\phi^+\rangle$}}
\put(14,24){\makebox(10,12){$S$}}
\put(14,44){\makebox(10,12){$R$}}
\put(14,74){\makebox(10,12){$B$}}
\put(14,94){\makebox(10,12){$A$}}

\put(30,30){\line(1,0){60}}
\put(30,50){\line(1,0){60}}
\put(30,80){\line(1,0){60}}
\put(30,100){\line(1,0){60}}

\put(50,100){\circle*{6}}
\put(50,100){\line(0,-1){55}}
\put(50,50){\circle{10}}

\put(70,80){\circle*{6}}
\put(70,80){\line(0,-1){55}}
\put(70,30){\circle{10}}

\put(140,34){\makebox(10,12){$|\phi^+\rangle$}}
\put(154,24){\makebox(10,12){$S'$}}
\put(154,44){\makebox(10,12){$R'$}}
\put(154,74){\makebox(10,12){$B$}}
\put(154,94){\makebox(10,12){$A$}}

\put(170,30){\line(1,0){60}}
\put(170,50){\line(1,0){60}}
\put(170,80){\line(1,0){60}}
\put(170,100){\line(1,0){60}}

\put(190,50){\circle*{6}}
\put(190,105){\line(0,-1){55}}
\put(190,100){\circle{10}}

\put(210,30){\circle*{6}}
\put(210,85){\line(0,-1){55}}
\put(210,80){\circle{10}}

\end{picture}
\vskip 0.1in
\caption{Local implementation of the Bell measurement superoperator using shared entanglement. In $(a)$, the two controlled-NOT gates imprint the parity bit of $AB$ onto the ancilla $RS$. In $(b)$, the two controlled-NOT gates imprint the phase bit of $AB$ onto the ancilla $R'S'$. Tracing over the ancillas in the Bell basis, we find that the $AB$ pair decoheres in the Bell basis.}
\label{fig:bell_circuit}
\end{figure}
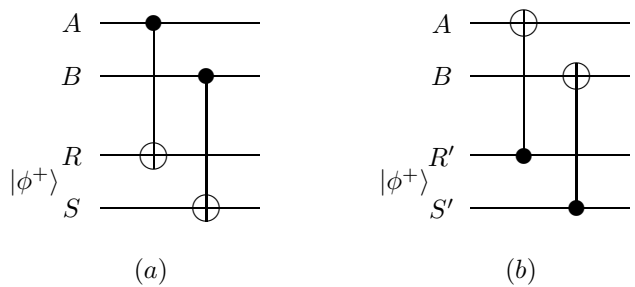

We note in passing that if Alice and Bob have ancilla pairs prepared in the state $|\phi^+\rangle_{RS}$ (where $R$ denotes Alice's ancilla qubit, and $S$ denotes Bob's), then they can implement the Bell measurement superoperator by executing the quantum circuits shown in Fig.~\ref{fig:bell_circuit}, and discarding their ancilla qubits. The circuit in Fig.~\ref{fig:bell_circuit}a flips the ancilla pair from $|\phi^+\rangle$ to $|\psi^+\rangle$ if the parity bit of the $AB$ state is $Z_A\otimes Z_B=-1$, and the circuit in Fig.~\ref{fig:bell_circuit}b flips the ancilla pair from $|\phi^+\rangle$ to $|\phi^-\rangle$ if the phase bit of the $AB$ state is $X_A\otimes X_B=-1$. Thus, the values of the parity and phase bits of the $AB$ pair, and only this information, become imprinted on the ancilla pairs. Tracing over the ancilla pairs in the Bell basis, then, induces decoherence in the $AB$ Bell basis.

In fact, if Alice and Bob share entanglement and perform the circuits of Fig.~\ref{fig:bell_circuit}, they can execute Bell measurement, including postselection, on the $AB$ pair by measuring their ancilla qubits and broadcasting their results classically. After executing circuit Fig.~\ref{fig:bell_circuit}a, Alice measures $Z_R$ and Bob measures $Z_S$. Then the parity bit of the $AB$ pair is the parity of these measurement outcomes. After executing circuit Fig.~\ref{fig:bell_circuit}b, Alice measures $X_{R'}$ and Bob measures $X_{S'}$. Then the phase bit of the $AB$ pair is the parity of the measurement outcomes. This example is instructive, as it reminds us again that entanglement is a more powerful resource than shared randomness. If Alice and Bob were limited to shared randomness and classical communication, they would be unable to create entanglement, and so would lack the capability of doing Bell measurement with postselection on their shared qubit pair.

\begin{figure}[h]
\centering
\begin{picture}(248,106)

\put(44,0){\makebox(10,12){($a$)}}
\put(184,0){\makebox(10,12){($b$)}}

\put(0,24){\makebox(10,12){$|0\rangle_S$}}
\put(0,44){\makebox(10,12){$|0\rangle_R$}}
\put(0,74){\makebox(10,12){$B$}}
\put(0,94){\makebox(10,12){$A$}}

\put(16,30){\line(1,0){60}}
\put(16,50){\line(1,0){60}}
\put(16,80){\line(1,0){60}}
\put(16,100){\line(1,0){60}}

\put(36,100){\circle*{6}}
\put(36,100){\line(0,-1){55}}
\put(36,50){\circle{10}}

\put(56,80){\circle*{6}}
\put(56,80){\line(0,-1){55}}
\put(56,30){\circle{10}}

\put(76,28){\framebox(24,24){${\rm Bell}\atop{\rm Meas.}$}}


\put(138,24){\makebox(10,12){$|+\rangle_{S'}$}}
\put(138,44){\makebox(10,12){$|+\rangle_{R'}$}}
\put(140,74){\makebox(10,12){$B$}}
\put(140,94){\makebox(10,12){$A$}}

\put(156,30){\line(1,0){60}}
\put(156,50){\line(1,0){60}}
\put(156,80){\line(1,0){60}}
\put(156,100){\line(1,0){60}}

\put(176,50){\circle*{6}}
\put(176,105){\line(0,-1){55}}
\put(176,100){\circle{10}}

\put(196,30){\circle*{6}}
\put(196,85){\line(0,-1){55}}
\put(196,80){\circle{10}}

\put(216,28){\framebox(24,24){${\rm Bell}\atop{\rm Meas.}$}}

\end{picture}
\vskip 0.1in
\caption{Bell measurement through entanglement swapping. Alice performs local controlled-NOT gates on her qubit $A$ and the ancilla qubits $RR'$. Bob performs local controlled-NOT gates on his qubit $B$ and the ancilla qubits $SS'$. Later, the ancilla qubits are collected, and Bell measurement is performed on the pairs $RS$ and $R'S'$. The Bell measurements on the ancilla realize Bell measurement on $AB$, by ``swapping'' entanglement of $AR$, $AR'$, $BS$, $BS'$ for entanglement of $AB$, $RR'$ and $SS'$. }
\label{fig:swap}
\end{figure}
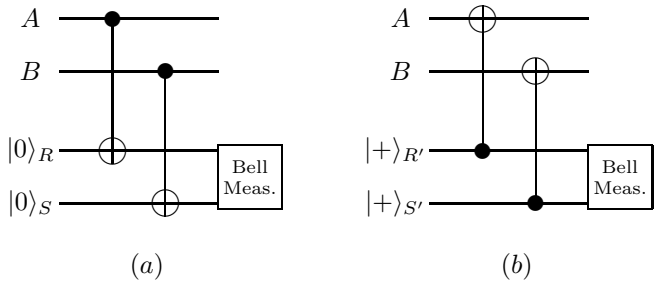

If Alice and Bob did not have entanglement to start with, they would still be able to perform Bell measurement with postselection on their shared pair if they could send ancilla qubits to a central laboratory for later quantum processing, as illustrated in Fig.~\ref{fig:swap}. Here Alice entangles her qubit $A$ first with ancilla qubit $R$ and then with $R'$, while Bob entangles his qubit $B$ first with ancilla qubit $S$ and then with $S'$. The ancilla qubits are collected, and a Bell measurement is performed on $RR'$ and $SS'$. The $RR'$ measurement yields the correct parity bit of the $AB$ pair, and a random phase bit; if the measured $RR'$ phase bit is $-1$, then the phase bit of the $AB$ state is flipped. The $SS'$ measurement yields the ``correct'' phase bit (possibly flipped by the $RR'$ measurement), and a random parity bit; if the measured $RR'$ parity bit is $-1$, then the parity bit of the $AB$ state is flipped. Taken together, the two Bell measurements on the ancilla qubits act as a projective measurement of ${\cal H}_{AB}$ onto the Bell basis, followed by one of the transformations $I\otimes I$, $X\otimes X$, $Y\otimes Y$, or $Z\otimes Z$; which of the four transformations has been applied, as well as the identity of the $AB$ Bell state resulting from the projection, can be inferred from outcomes of the $RR'$ and $SS'$ measurements.
If the $AB$ state is initially a product state, Bell measurement of the ancilla qubits establishes entanglement of $A$ with $B$ by ``swapping'' $AR$ and $BS$ entanglement for $AB$ and $RS$ entanglement \cite{zukowski}.

This example is also instructive. It reminds us that decoherence induced on a time slice can sometimes be reversed later through the operation of a ``quantum eraser'' \cite{scully}. If we were to trace out the ancilla qubits right after applying the controlled-NOT's of Fig.~\ref{fig:swap} (before the Bell measurement), then Fig~\ref{fig:swap}a would induce decoherence, not in an entangled basis, but rather in the product basis $\{|00\rangle, |01\rangle, |10\rangle,|11\rangle\}$; similarly Fig~\ref{fig:swap}b would induce decoherence in the product basis $\{|++\rangle, |+-\rangle, |-+\rangle,|--\rangle\}$, where
\begin{equation}
|\pm\rangle= {1\over \sqrt{2}}\left(|0\rangle \pm |1\rangle\right)~.
\end{equation}
As one would expect, without entanglement or shared randomness, Alice and Bob are unable to implement decoherence in the Bell basis with their local operations alone. But a later measurement {\em including postselection} can ``undo'' the decoherence in the product basis and establish decoherence in the Bell basis instead. 

\subsection{Local stabilizers}
The observation that shared randomness is sufficient to induce decoherence in the Bell basis can be substantially generalized. Consider a superoperator ${\cal E}$ that acts on a density operator $\rho$ as
\begin{equation}
\label{group_so}
{\cal E}(\rho)= {1\over |G|}\sum_{g\in G} U(g)~\rho ~U(g)^\dagger~,
\end{equation}
where the $U(g)$'s provide a (not necessarily irreducible) unitary representation of the group $G$, and $|G|$ denotes the order of $G$. The Hilbert space ${\cal H}$ in which $\rho$ resides can be decomposed into spaces that transform irreducibly under the group $G$. Let us choose an orthonormal basis
\begin{equation}
\{|R,a,i\rangle\}~;
\end{equation}
here $R$ labels the irreducible representations of $G$, $a$ labels the sectors of ${\cal H}$ that transform as the irreducible representation $R$ (a particular irreducible representation might occur multiple times), $i=1,2,\dots, n_R$ labels states of a basis for the vector space on which $R$ acts, and $n_R$ is the dimension of $R$. Expressed in this basis, the representation $U(g)$ is
\begin{equation}
\label{rep_in_basis}
U(g)=\sum_{R,a,i,j} |R,a,i\rangle ~D^{(R)}(g)_{ij}~\langle R,a,j|~,
\end{equation}
where $D^{(R)}(g)_{ij}$ is a matrix element of the irreducible representation $R$. These matrix elements obey the orthogonality relations
\begin{equation}
{1\over |G|}\sum_{g\in G} D^{(R)}(g)_{ij} D^{(R')}(g)_{kl}^*= {1\over n_R}\delta^{RR'}\delta_{ik}\delta_{jl}~.
\end{equation}
Substituting eq.~(\ref{rep_in_basis}) into eq.~(\ref{group_so}) and applying the orthogonality relations, we find
\begin{eqnarray}
\label{group_decohere}
&&\langle R,a,i|{\cal E}(\rho)|R',b,j\rangle \nonumber\\
&& =\delta^{RR'}\cdot {1\over n_R}\delta_{ij}\cdot
\sum_{k=1}^{n_R}\langle R,a,k|\rho|R,b,k\rangle~.
\end{eqnarray}
Thus we see that the superoperator ${\cal E}$ destroys the coherence of a superposition of states that transform as distinct irreducible representations of $G$. Within a given irreducible representation, it randomizes the state, replacing the density operator by a multiple of the identity. Some off-diagonal terms in the density operator can survive, if a given irreducible representation occurs in ${\cal H}$ more than once.

Now suppose that the Hilbert space ${\cal H}$ is shared by $n$ parties; it has a tensor product decomposition
\begin{equation}
{\cal H}=\otimes_{\alpha=1}^n {\cal H}_{\alpha}~.
\end{equation}
And suppose that each element $U(g)$ of the representation of $G$ is a tensor product
\begin{equation}
U(g)=\otimes_{\alpha=1}^n U(g)_{\alpha}~.
\end{equation}
Then the $n$ parties can perform the operation ${\cal E}$ by consulting their shared randomness -- if they are instructed to apply the group element $g\in G$, the party $\alpha$ applies $U(g)_\alpha$ to her portion of the state.

In the Bell measurement case discussed above, the four-dimensional Hilbert space of two qubits transforms as the representation 
\begin{equation}
\{I\otimes I, X\otimes X, - Y\otimes Y, Z\otimes Z\}~.
\end{equation}
of the group $G=Z_2\times Z_2$. The group $G$ is Abelian, and the four Bell states transform as distinct one-dimensional irreducible representations of $G$. Therefore, the superoperator ${\cal E}$ induces decoherence in the Bell basis.  

The same ideas apply to any {\em stabilizer code} \cite{gottesman,att,nielsen_chuang,preskill229}. Consider, for example, an Abelian group $G$ acting on $n$ qubits, generated by $n-k$ operators, where each generator is a tensor product of single-qubit Pauli operators. If each qubit is entrusted to a distinct party, then by consulting shared randomness, the $n$ parties can apply a random element of the group $G$ to their density operator. The superoperator they implement acts trivially on each $2^k$-dimensional code space -- an eigenspace of the generators with specified eigenvalues. But it destroys the coherence of a superposition of eigenspaces with different eigenvalues. In the notation of eq.~(\ref{group_decohere}), the index $R$ labels the stabilizer eigenvalues, and the indices $a,b$ label the basis states in a code space with a specified value of $R$. Because the group is Abelian, all of the irreducible representations are one-dimensional, and eq.~(\ref{group_decohere}) becomes
\begin{eqnarray}
{\cal E}(\rho)&=&{1\over |G|}\sum_{g\in G} U(g)~\rho ~U(g)^\dagger\nonumber\\
&=& \sum_{R,a,b}|R,a\rangle\langle R,a|\rho|R,b\rangle\langle R,b|= \sum_R E_R \rho E_R~,
\end{eqnarray}
where $E_R$ projects onto the subspace with specified stabilizer eigenvalues.

The observation that Bell {\em measurement} (including postselection) can be achieved with shared entanglement, local operations, and broadcasting of classical data can be generalized to any CSS stabilizer code; {\it i.e.}, any code of the class constructed by Calderbank and Shor \cite{cal_shor}, and Steane \cite{steane_code}. An $n$-qubit stabilizer code is of the CSS type if the stabilizer generators can be chosen so that each generator is either a tensor product of $Z$'s and $I$'s or a tensor product of $X$'s and $I$'s.  Imagine that each of the $n$ qubits is in the custody of a separate party.  Steane \cite{steane} has observed that the measurement of the stabilizer generators can always be achieved by carrying out these steps: (1) preparation of a suitable entangled ancilla that is distributed to the $n$ parties, (2) local quantum gates applied by each party, acting on her qubit and her part of the ancilla, (3) local measurements by each party, and (4) classical post-processing of the measurement outcomes. In the case of Bell measurement on a pair of qubits, the stabilizer generators are $X\otimes X$, $Z\otimes Z$, and the entangled ancilla state is $|\phi^+\rangle$.

An example of a superoperator associated with a non-Abelian group is the ``twirling'' operation that transforms a two-qubit state into a Werner state \cite{bdsw}. In that case, the group is $A_4$, the order-12 subgroup of the rotation group that preserves a tetrahedron. Under $A_4$, the state $|\psi^-\rangle$ transforms trivially, while the other three Bell states $|\phi^+\rangle, |\psi^+\rangle, |\phi^-\rangle$ transform as a three-dimensional irreducible representation. Two parties Alice and Bob consult their shared randomness and apply a random element of $A_4$; according to eq.~(\ref{group_decohere}), this operation transforms any initial density operator into a state of the Werner form
\begin{eqnarray}
&&\rho(F)= F|\psi^-\rangle\langle \psi^-| \nonumber\\
&&+ {1-F\over 3}\left(|\phi^+\rangle\langle \phi^+|+|\phi^-\rangle\langle \phi^-|+|\psi^+\rangle\langle \psi^+|\right)~,
\end{eqnarray}
while preserving the fidelity $F=\langle\psi^-|\rho|\psi^-\rangle$. Note that unlike the acausal operation defined by eq.~(\ref{sorkin}), this localizable operation transforms any initial state into an incoherent mixture of Bell states; hence Alice's final density operator is always $I/2$, and Alice is unable to receive a signal from Bob.

\section{Causal and semicausal complete measurement superoperators}

In the next three sections, we will investigate the causality and localizability properties of measurement superoperators that project onto a complete orthonormal basis. We will show that semicausal operations of this class are semilocalizable, and that fully causal operations of this class are not necessarily fully localizable.

Suppose that the Hilbert spaces ${\cal H}_A$ and ${\cal H}_B$ have dimensions $N_A$ and $N_B$ respectively, and let $\{E^a = |a\rangle\langle a|, a = 1,2, \dots, N_A N_B\}$ denote a complete set of orthogonal one-dimensional projectors on ${\cal H}_A\otimes{\cal H}_B$. By tracing over ${\cal H}_B$, we obtain from these projectors $N_AN_B$ positive operators acting on ${\cal H}_A$, each with unit trace, defined by 
\begin{equation}
\sigma^a_A={\rm tr}_B\left(E^a_{AB}\right)~.
\end{equation}
Since the $E^a_{AB}$'s are complete, these operators satisfy
\begin{equation}
\label{sigma_sum}
\sum_a \sigma_A^a = {\rm tr}_B~I_{AB} = N_B~I_A~;
\end{equation}
that is, $\{{N_B}^{-1}~\sigma_A^a\}$ is a positive operator-valued measure (POVM) on ${\cal H}_A$ with $N_AN_B$ outcomes.

The semicausal complete orthogonal measurement superoperators (those that do not allow Bob to signal Alice) can be simply characterized by a property of the $\sigma_A^a$'s, thanks to the following theorem:

\noindent {\bf Theorem 3} {\em A complete orthogonal measurement superoperator is semicausal if and only if it has the following property: For each pair of operators $\{\sigma_A^a,\sigma_A^b\}$, either $\sigma_A^a=\sigma_A^b$ or $\sigma_A^a\sigma_A^b=0=\sigma_A^b\sigma_A^a$.}

\noindent That is, any pair of $\sigma_A^a$'s must be either identical or orthogonal, if and only if the superoperator is semicausal. 

Theorem 3 is proved in Appendix A, but one can readily see that the result is plausible. If $E^a = |a\rangle\langle a|$, then, in order to signal Alice, Bob by acting locally needs to induce a transition from the state $|a\rangle$ to the state $|b\rangle$ for some $a$ and $b$; furthermore, Alice must be able to detect the difference between $|a\rangle$ and $|b\rangle$. But if $\sigma^a_A$ and $\sigma^b_A$ are orthogonal, then Bob is unable to induce the transition, and if $\sigma^a_A=\sigma^b_A$, then Alice can't tell the difference. On the other hand, if $\sigma^a_A\sigma^b_A\ne 0$ and $\sigma^a_A\ne \sigma^b_A$, then Bob can induce the transition and Alice can distinguish the states; hence a signaling protocol can be devised. 

Applying Theorem 3, we can see that all semicausal complete measurements have a simple structure. Suppose that $\{E^a\}$ is a semicausal complete measurement. For any one of the $\sigma_A^a$, let  ${\cal H}_A^a$ denote the subspace of ${\cal H}_A$ on which $\sigma_A^a$ has its support ($\sigma_A^a$ is strictly positive on this subspace and vanishes on the orthogonal subspace). According to Theorem 3, the support ${\cal H}_A^b$ of each $\sigma_A^b$ either coincides with ${\cal H}_A^a$ or is orthogonal to ${\cal H}_A^a$, and in the former case we have $\sigma_A^a=\sigma_A^b$.

Thus each $\sigma_A^b$ with support on ${\cal H}_A^a$ must equal $\sigma_A^a$, and furthermore, the sum of the operators with support on this subspace, according to eq.~(\ref{sigma_sum}), must be $N_B\cdot I_A^a$, where $I_A^a$ denotes the projector onto ${\cal H}_A^a$.  Therefore each $\sigma_A^a$ with support on ${\cal H}_A^a$ is proportional to $I_A^a$, and since we also know that ${\rm tr}~\sigma_A^a=1$, must be
\begin{equation}
\sigma_A^a= {1\over d_A^a}\cdot I_A^a~,
\end{equation}
where $d_A^a$ is the dimension of ${\cal H}_A^a$. We also conclude that the number of $\sigma_A^b$'s with support on ${\cal H}_A^a$ is $N_B\cdot d_A^a$.

\begin{figure}[h]
\centering
\begin{picture}(140,146)

\put(74,134){\makebox(12,12){Bob}}
\put(0,64){\makebox(12,12){Alice}}
\put(20,10){\line(1,0){120}}
\put(20,30){\line(1,0){120}}
\put(74,14){\makebox(12,12){$1\times 6$}}

\put(20,10){\line(0,1){120}}
\put(20,130){\line(1,0){120}}
\put(140,130){\line(0,-1){120}}

\put(20,70){\line(1,0){120}}
\put(74,44){\makebox(12,12){$2\times 6$}}
\put(74,94){\makebox(12,12){$3\times 6$}}
\end{picture}
\caption{A semicausal complete orthogonal measurement in $6\times 6$ dimensions. Alice's Hilbert space is decomposed into three mutually orthogonal subspaces, of dimensions 3, 2, and 1. The measurement projects onto an orthonormal basis, where each element of the basis is a maximally entangled state of one of Alice's three subspaces with Bob's space.}
\label{fig:semicausal}
\end{figure}
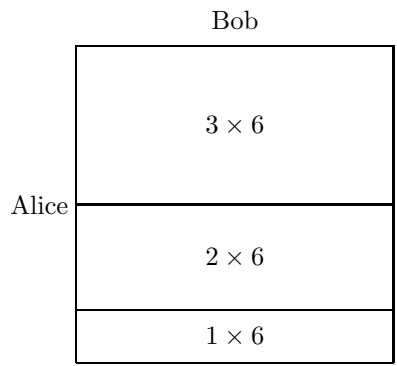

We see that the state $|a\rangle$ is a maximally entangled state of the form
\begin{equation}
|a\rangle= {1\over \sqrt{d_A^a}}\sum_{i=1}^{d_A^a}|i\rangle_A\otimes |i'\rangle_B~,
\end{equation}
where here $\{|i\rangle_A\}$ denotes an orthonormal basis for ${\cal H}_A^a$, and the $|i'\rangle_B$'s are mutually orthogonal states of ${\cal H}_B$. The general structure of a semicausal complete measurement operation, then, is as illustrated in Fig~\ref{fig:semicausal}.  Alice's Hilbert space can be decomposed into mutually orthogonal subspaces
\begin{equation}
{\cal H}_A=\oplus_\alpha {\cal H}_A^\alpha~,
\end{equation}
where ${\cal H}_A^\alpha$ has dimension $d_A^\alpha$. Of the $N_A\cdot N_B$ states comprising the basis $\{|a\rangle\}$, $N_B\cdot d_A^\alpha$, all maximally entangled, have support on ${\cal H}_A^\alpha$.  

A {\em fully causal} complete measurement is more highly constrained.  Since the measurement is semicausal in both directions, both the $\sigma_A^a$'s and the $\sigma_B^a$'s obey the conditions specified in Theorem 3. If we choose one particular ${\cal H}_A^\alpha$, there are $N_B\cdot d_A^\alpha$ elements of the basis with support on this space, and associated with these are $N_B\cdot d_A^\alpha$ $\sigma_B^a$'s, all of rank $d_A^\alpha$, and any two of which must either coincide or be orthogonal.  Therefore, the $\sigma_B^a$'s partition ${\cal H}_B$ into mutually orthogonal subspaces, all of dimension $d_A^\alpha$; it follows that $d_A^\alpha$ must divide $N_B$, that the number of orthogonal subspaces is $N_B/d_A^\alpha$, and that $(d_A^\alpha)^2$ $\sigma_B^a$'s have support on each space.

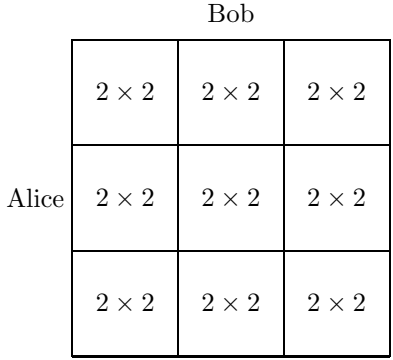
\begin{figure}[h]
\centering
\begin{picture}(140,146)

\put(74,134){\makebox(12,12){Bob}}
\put(0,64){\makebox(12,12){Alice}}

\put(20,10){\line(1,0){120}}
\put(20,10){\line(0,1){120}}
\put(20,130){\line(1,0){120}}
\put(140,130){\line(0,-1){120}}

\put(20,50){\line(1,0){120}}
\put(20,90){\line(1,0){120}}

\put(60,10){\line(0,1){120}}
\put(100,10){\line(0,1){120}}

\put(34,24){\makebox(12,12){$2\times 2$}}
\put(74,24){\makebox(12,12){$2\times 2$}}
\put(114,24){\makebox(12,12){$2\times 2$}}

\put(34,64){\makebox(12,12){$2\times 2$}}
\put(74,64){\makebox(12,12){$2\times 2$}}
\put(114,64){\makebox(12,12){$2\times 2$}}

\put(34,104){\makebox(12,12){$2\times 2$}}
\put(74,104){\makebox(12,12){$2\times 2$}}
\put(114,104){\makebox(12,12){$2\times 2$}}

\end{picture}
\caption{A causal complete orthogonal measurement in $6\times 6$ dimensions. Alice's 6-dimensional Hilbert space is partitioned into 3 mutually orthogonal subspaces, each of dimension 2, and Bob's Hilbert space is similarly partitioned. The measurement projects onto an orthonormal basis, where each element of the basis is a maximally entangled state of one of Alice's subspaces with one of Bob's subspaces.}
\label{fig:causal}
\end{figure}

Applying the same argument again, but with ${\cal H}_A$ and ${\cal H}_B$ interchanged, we see that the $\sigma_A^a$'s also partition ${\cal H}_A$ into mutually orthogonal subspaces, all of dimension $d_A^\alpha$. We conclude that a causal complete measurement has the structure illustrated in Fig.~\ref{fig:causal}. Alice's space can be decomposed into $r_A$ subspaces ${\cal H}_A^\alpha$, each of dimension $d$ (so that $N_A= r_A\cdot d$), and Bob's space can be decomposed into $r_B$ subspaces ${\cal H}_B^\beta$, each of dimension $d$ (with $N_B=r_B\cdot d$). The measurement projects onto an orthonormal basis, where each element of the basis is a maximally entangled state of some ${\cal H}_A^\alpha$ with some ${\cal H}_B^\beta$. There are $r_A\cdot r_B$ ways to choose $\alpha$ and $\beta$, and there are $d^2$ maximally entangled states for each pair of subspaces.

The extreme cases are $d=1$, for which we have a product basis $\{|\alpha,\beta\rangle_{AB}\}$, and $d=N_A=N_B$, for which the measurement is a projection onto a maximally entangled basis of ${\cal H}_{AB}$.

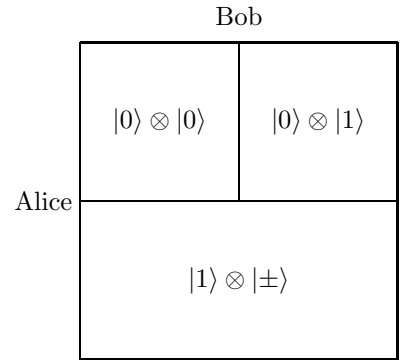
\begin{figure}[h]
\centering
\begin{picture}(140,146)

\put(74,134){\makebox(12,12){Bob}}
\put(0,64){\makebox(12,12){Alice}}

\put(20,10){\line(1,0){120}}
\put(20,10){\line(0,1){120}}
\put(20,130){\line(1,0){120}}
\put(140,130){\line(0,-1){120}}

\put(20,70){\line(1,0){120}}
\put(80,70){\line(0,1){60}}

\put(44,94){\makebox(12,12){$|0\rangle\otimes |0\rangle$}}
\put(104,94){\makebox(12,12){$|0\rangle\otimes |1\rangle$}}
\put(74,34){\makebox(12,12){$|1\rangle\otimes |\pm\rangle$}}

\end{picture}
\caption{A semicausal complete orthogonal measurement in $2\times 2$ dimensions. The orthonormal basis shown partitions Alice's space into mutually orthogonal one-dimensional subspaces; hence Bob cannot signal Alice. But since Bob's space is not so partitioned, Alice can signal Bob.}
\label{fig:2by2-causal}
\end{figure}

Comparing Fig.~\ref{fig:semicausal} and Fig.~\ref{fig:causal} makes it clear that a semicausal measurement need not be fully causal. Indeed, this feature is quite obvious, since transmission of information from Alice to Bob can allow Alice to signal Bob but does not enable Bob to signal Alice. To make this point more explicit, consider the $2\times 2$ example illustrated in Fig.~\ref{fig:2by2-causal}. The measurement projects onto the orthonormal basis 
\begin{eqnarray}
\label{causal_not_semicausal}
& |0\rangle_A\otimes |0\rangle_B~,\quad &|0\rangle_A\otimes |1\rangle_B~,\nonumber\\
& |1\rangle_A\otimes |+\rangle_B~,\quad &|1\rangle_A\otimes |-\rangle_B~,
\end{eqnarray}
where 
\begin{equation}
|\pm\rangle= {1\over\sqrt{2}}\left(|0\rangle\pm |1\rangle\right)~.
\end{equation}
Tracing over Bob's system we obtain Alice's projectors
\begin{eqnarray}
\sigma_A^{00}&=\sigma_A^{01}= &|0\rangle\langle 0|~, \nonumber\\
\sigma_A^{1+}&=\sigma_A^{1-}= &|1\rangle \langle 1|~,
\end{eqnarray}
which satisfy the criterion of Theorem 3; hence Bob cannot signal Alice. Tracing over Alice's system we obtain Bob's projectors
\begin{eqnarray}
\sigma_B^{00}&=|0\rangle\langle 0|~,\quad \sigma_B^{01}&= |1\rangle\langle 1|~, \nonumber\\
\sigma_B^{1+}&=|+\rangle\langle +|~,\quad  \sigma_B^{1-}&= |-\rangle \langle -|~,
\end{eqnarray}
which violate the criterion of Theorem 3; hence Alice can signal Bob.

A particular protocol that allows Alice to signal Bob works as follows: Bob prepares his qubit in the state $|0\rangle$ and Alice prepares hers in one of the states $|0\rangle$, $|1\rangle$. After the operation ${\cal E}$ is applied, Bob's density operator is
\begin{equation}
\rho^0_B= |0\rangle\langle 0|~,
\end{equation}
if Alice prepared $|0\rangle$, and his density operator is 
\begin{equation}
\rho^1_B= I/2~,
\end{equation}
if Alice prepared $|1\rangle$. Since $\rho_B^0\ne \rho_B^1$, we conclude that ${\cal E}$ is not causal.

Note that the measurement that projects onto the basis eq.~(\ref{causal_not_semicausal}) is obviously semilocalizable; in fact it can be executed with one-way {\em classical} communication from Alice to Bob. Alice measures in the basis $\{|0\rangle_A,|1\rangle_A\}$, and sends her measurement outcome to Bob. Then Bob measures in the basis $\{|0\rangle_B,|1\rangle_B\}$ if Alice's outcome was $|0\rangle_A$, and in the basis $\{|+\rangle_B,|-\rangle_B\}$ if Alice's outcome was $|1\rangle_A$.

\section{semicausal complete measurement superoperators are semilocalizable}
We now have learned enough about the structure of semicausal complete
measurements to see that any semicausal complete measurement 
superoperator is semilocalizable.  Suppose that Alice and Bob share a quantum state, and wish to perform a measurement that projects onto the basis $\{|a\rangle\}$, where each $|a\rangle$
is a maximally entangled state of a subspace of ${\cal H}_A$ with
a subspace of ${\cal H}_B$. Alice can perform a partial measurement that identifies the subspace ${\cal H}_A^\alpha$ of ${\cal H}_A$, and then send her half of the state to Bob, who can finish the measurement and identify $|a\rangle$. To complete the procedure, they simply replace their original state by a state identical to $|a\rangle$ that can be drawn from their shared ancilla. Since Bob can convert any maximally entangled state of ${\cal H}_A^\alpha$ and ${\cal H}_B$ into $|a\rangle$ by performing a unitary transformation on ${\cal H}_B$, he and Alice can replace the original state by $|a\rangle$ without any further communication. 
From these observations we obtain:

\noindent {\bf Theorem 4} {\em A semicausal complete orthogonal measurement superoperator is semilocalizable}.

\noindent{\bf Proof}: Suppose that Alice and Bob share the state $\rho_{AB}$ in their joint Hilbert space ${\cal H}_A\otimes {\cal H}_B$.  In addition, an ancillary Hilbert space ${\cal H}_R\otimes {\cal H}_S$, isomorphic to ${\cal H}_A\otimes {\cal H}_B$, is initially under Alice's control. To implement the semicausal operation 
\begin{equation}
{\cal E}_{AB}(\rho_{AB})=\sum_a E_{AB}^a\left(\rho_{AB}\right) E_{AB}^a
\end{equation}
with one-way quantum communication from Alice to Bob, they proceed as follows: First, Alice performs a partial measurement that projects $\rho_{AB}$ onto her mutually orthogonal subspaces ${\cal H}_A^\alpha$, obtaining the outcome
\begin{equation}
\rho_{AB}^\alpha={E_A^\alpha \left(\rho_{AB}\right) E_A^\alpha \over {\rm tr}\left(E_A^\alpha\rho_{AB}\right)}
\end{equation}
with probability 
\begin{equation}
p_\alpha = {\rm tr} \left(E_A^\alpha\rho_{AB}\right)~,
\end{equation}
where $E_A^\alpha$ is the projector onto ${\cal H}_A^\alpha$. If Alice's measurement outcome is $\alpha$, she prepares an ancilla state $|\Phi\rangle_{RS}\in {\cal H}_R\otimes {\cal H}_S$, and a dimension $d_A^\alpha \times d_A^\alpha$ maximally entangled state of ${\cal H}_R^\alpha$ with ${\cal H}_S$, where ${\cal H}_R^\alpha$ is isomorphic to ${\cal H}_A^\alpha$.  Next, Alice swaps the Hilbert spaces ${\cal H}_A$ and ${\cal H}_R$, obtaining $\rho_{RB}^\alpha$ and $|\Phi\rangle_{AS}$. She sends $\rho_{RB}^\alpha$ to Bob, along with the $S$ half of the entangled state $|\Phi\rangle_{AS}$.

Upon receipt, Bob swaps the Hilbert spaces ${\cal H}_B$ and ${\cal H}_S$, so that Alice and Bob now share $|\Phi\rangle_{AB}$, while $\rho_{RS}^\alpha$ is entirely in Bob's hands. On the state $\rho_{RS}^\alpha$, Bob performs an orthogonal measurement with projectors $E_{RS}^a$ that are isomorphic to the $E_{AB}^a$'s, obtaining the outcome $|a\rangle_{RS}$ with probability $p_{a|\alpha}$, where
\begin{equation}
p_{a,\alpha}= p_{a|\alpha}\cdot p_\alpha = {\rm tr} \left(E_{RS}^aE_{R}^\alpha\rho_{RS} E_R^\alpha E_{RS}^a\right)~.
\end{equation}
Since Bob's measurement is just a completion of Alice's partial measurement, we have
\begin{equation}
E_{RS}^a E_R^\alpha = E_{R}^\alpha E_{RS}^a= \delta_{a,\alpha} \cdot E_{RS}^a~,
\end{equation}
where $\delta_{a,\alpha}$ is 1 if $\sigma_R^a$ has support on ${\cal H}_R^\alpha$ and 0 otherwise. Therefore, the probability that Bob obtains outcome $a$ can be expressed as
\begin{eqnarray}
p_a=\sum_\alpha p_{a,\alpha}&= &{\rm tr}_{RS} \left(E_{RS}^a\rho_{RS}\right)\nonumber\\
&=& {\rm tr}_{AB} \left(E_{AB}^a\rho_{AB}\right)~.
\end{eqnarray}
(Bob's measurement commutes with Alice's, so it is just as though Bob measured first, and Alice has been provided with incomplete information about what Bob found.)

Now, since the state $|a\rangle_{RS}$ prepared by Bob's measurement is a $d_A^\alpha\times d_A^\alpha$ maximally entangled state of ${\cal H}_R^\alpha$ with ${\cal H}_S$, Bob can apply a suitable unitary transformation to his half of the state $|\Phi\rangle_{AB}$ that he now shares with Alice, rotating it to the state $|a\rangle_{AB}$. Thus, Alice and Bob have converted their initial state $\rho_{AB}$ to $|a\rangle_{AB}$ with probability $p_a=\langle a|\rho_{AB}|a\rangle$. Finally, Bob discards the ancilla $RS$, and Alice and Bob discard the record of their measurement outcomes. We have described a protocol with one-way quantum communication that executes the semicausal complete measurement superoperator ${\cal E}_{AB}$. This proves Theorem 4.

Note that if we dispense with the last step, in which Alice and Bob discard their records, then we see that not just the measurement superoperator, but also the measurement operation with postselection, is semilocalizable: with one way quantum communication from Alice to Bob, the state is projected onto the basis, and the measurement outcome is known by Bob (though not by Alice).

\section{Causal complete measurement superoperators need not be localizable}
Now that the general structure of causal complete measurements is known, we can address whether causal complete measurements are localizable. In fact, we will be able to construct examples of causal measurements that are provably {\em not} localizable. To accomplish this task, we will identify a property satisfied by localizable operations, and exhibit causal measurements that don't possess this property.

We say that a (not necessarily normalized) pure state $|\psi\rangle$ is an {\em eigenstate} of a superoperator ${\cal E}$ if
\begin{equation}
{\cal E}(|\psi\rangle\langle \psi|)= |\psi\rangle\langle \psi|~.
\end{equation}
The key property of localizable superoperators that we will exploit is:

\noindent {\bf Theorem 5} {\em If ${\cal E}$ is a localizable superoperator on ${\cal H}_A\otimes {\cal H}_B$, and $|\psi\rangle$, $A\otimes I|\psi\rangle$, and $I\otimes B|\psi\rangle$ are all eigenstates of ${\cal E}$ (where $A$ and $B$ are invertible operators), then $A\otimes B|\psi\rangle$ is also an eigenstate of ${\cal E}.$}

\noindent The proof of Theorem 5 is in Appendix B. Clearly the claim is plausible. By hypothesis, the eigenstate $|\psi\rangle$ of ${\cal E}$ is mapped to a new eigenstate if Alice applies $A$ and Bob does nothing, or if Bob applies $B$ and Alice does nothing. Since ${\cal E}$ is localizable, Alice and Bob should be able to decide independently whether to apply $A$ and $B$, and still obtain an eigenstate of ${\cal E}$. 

\subsection{A twisted partition}
Having identified in Theorem 5 a necessary condition for localizability, we proceed to describe causal measurements for which this condition is violated. An example in $4\times 4$ dimensions is illustrated in Fig.~\ref{fig:causnotloc}. Alice's four-dimensional space is partitioned into two two-dimensional subspaces, ${\cal H}_A^{01}$ spanned by $\{|0\rangle_A, |1\rangle_A\}$, and ${\cal H}_A^{23}$ spanned by $\{|2\rangle_A,|3\rangle_A\}$. Bob's space is similarly partitioned. Let's first consider the case where the measurement projects onto the standard Bell basis $\{|\phi^\pm\rangle,|\psi^\pm\rangle\}$ in each of the four subspaces ${\cal H}_A^\alpha \otimes {\cal H}_B^\beta$; that is, the orthonormal basis is
\begin{eqnarray}
&|\phi_{00}^\pm\rangle, |\psi_{00}^\pm\rangle,\quad &|\phi_{02}^\pm\rangle, |\psi_{02}^\pm\rangle~, \nonumber\\
&|\phi_{20}^\pm\rangle, |\psi_{20}^\pm\rangle,\quad &|\phi_{22}^\pm\rangle, |\psi_{22}^\pm\rangle~, 
\end{eqnarray}
where we use the notation
\begin{eqnarray}
&&|\phi_{ij}^\pm\rangle= {1\over\sqrt{2}}\left(|i,j\rangle \pm |i+1,j+1\rangle\right)~,\nonumber\\
&&|\psi_{ij}^\pm\rangle= {1\over\sqrt{2}}\left(|i,j+1\rangle \pm |i+1,j\rangle\right)~.
\end{eqnarray}

The superoperator that induces decoherence in this basis {\em is} localizable, and in fact it can be implemented with shared randomness -- no entanglement is required. Alice and Bob can each perform a partial measurement to identify whether the state occupies the subspace ${\cal H}^{01}$ or ${\cal H}^{23}$. Then they can proceed to implement decoherence in the $2\times 2$ Bell basis as described in \S IIC. Finally, Alice and Bob discard the record of the partial measurement to complete the implementation of the measurement superoperator.

But now consider a ``twisted'' basis in which the basis elements in the ${\cal H}_A^{23} \otimes {\cal H}_B^{23}$ quadrant of the Hilbert space are rotated by applying a unitary transformation $U_B$  to Bob's half of the state, becoming
\begin{equation}
I_A\otimes U_B|\phi_{22}^\pm\rangle~,\quad I_A\otimes U_B|\psi_{22}^\pm\rangle~,
\end{equation}
where $U_B$ maps ${\cal H}_B^{23}$ to ${\cal H}_B^{23}$.
Since this new basis still meets the criterion of Theorem 3 in both ${\cal H}_A$ and ${\cal H}_B$, the corresponding measurement superoperator ${\cal E}$ is still causal. But because ${\cal E}$ does not satisfy the criterion of Theorem 5 (except in the case were $U_B$ merely permutes the Bell basis), it is no longer localizable.  The eigenstates of ${\cal E}$ include, for example,
\begin{eqnarray}
&&|\phi_{00}^\pm\rangle~, \nonumber\\
&&|\phi_{20}^\pm\rangle=X^2\otimes I|\phi_{00}^\pm\rangle~, \nonumber\\
&&|\phi_{02}^\pm\rangle=I\otimes X^2|\phi_{00}^\pm\rangle~,
\end{eqnarray}
where $X^2$ is the four-dimensional Pauli operator that acts on the basis $\{|0\rangle,|1\rangle, |2\rangle, |3\rangle\}$ according to 
\begin{equation}
X^2: |i\rangle \to |i+2~{\rm (mod ~4)}\rangle~.
\end{equation}
If ${\cal E}$ is localizable, Theorem 5 requires that 
\begin{equation}
|\phi_{22}^\pm\rangle =X^2\otimes X^2|\phi^\pm\rangle
\end{equation}
also be an eigenstate -- {\it i.e.}, that $|\phi_{22}\rangle$ is also an element of the orthonormal basis. This isn't so unless $U_B$ is one of the Pauli matrices, up to a phase. Therefore, ${\cal E}$ is not localizable.

\begin{figure}[h]
\centering
\begin{picture}(210,96)

\put(114,84){\makebox(12,12){Bob}}
\put(0,34){\makebox(12,12){Alice}}
\put(69,74){\makebox(12,12){01}}
\put(159,74){\makebox(12,12){23}}
\put(14,19){\makebox(12,12){23}}
\put(14,49){\makebox(12,12){01}}
\put(30,10){\line(1,0){180}}
\put(30,10){\line(0,1){60}}
\put(30,70){\line(1,0){180}}
\put(210,10){\line(0,1){60}}

\put(30,40){\line(1,0){180}}
\put(120,10){\line(0,1){60}}

\put(69,49){\makebox(12,12){$|\phi^\pm\rangle, |\psi^\pm\rangle$}}
\put(69,19){\makebox(12,12){$|\phi^\pm\rangle, |\psi^\pm\rangle$}}
\put(158,49){\makebox(12,12){$|\phi^\pm\rangle, |\psi^\pm\rangle$}}
\put(158,19){\makebox(12,12){$U_B|\phi^\pm\rangle, U_B|\psi^\pm\rangle$}}

\end{picture}
\caption{A causal complete orthogonal measurement that is not localizable, in $4\times 4$ dimensions. The Hilbert space is divided into four $2\times 2$ quadrants, and the elements of the orthonormal basis are maximally entangled Bell states in each quadrant. Because the Bell basis in the bottom right quadrant has been twisted by applying the unitary transformation $I\otimes U_B$, the corresponding measurement superoperator cannot be implemented without communication between Alice and Bob.}
\label{fig:causnotloc}
\end{figure}

The method that worked for the untwisted basis ($U_B=I$) fails for the twisted basis. Alice's partial measurement identifies what row the state occupies, and Bob's measurement identifies the column, but neither one has enough information to determine whether or not the state lies in the bottom right quadrant where the basis is twisted. Without this information, they can't complete the protocol successfully.

If Bob did have this information, then the protocol could be completed. Hence, not only is ${\cal E}$ semilocalizable (like any causal complete measurement superoperator); furthermore it can actually be implemented with one-way {\em classical} communication. Alice performs the partial measurement that projects onto ${\cal H}_A^{01}$ or ${\cal H}_A^{23}$, and sends her measurement outcome to Bob. She also sends to Bob a copy of a table of random numbers that she has generated. Then Bob, after performing his partial measurement, has enough information to determine whether the state occupies the bottom right quadrant, where the stabilizer generators are
\begin{equation}
X\otimes U_B X U_B^{-1}~,\quad Z\otimes U_B Z U_B^{-1}~,
\label{twisted_bell_stab}
\end{equation}
or one of the other three quadrants, where the stabilizer generators are
\begin{equation}
X\otimes X~,\quad Z\otimes Z~.
\label{bell_stab}
\end{equation}
(Here $X$ and $Z$ denote $2\times 2$ Pauli operators.)
When Alice and Bob consult their shared randomness and Alice is directed to apply $\sigma\in \{I,X,Y,Z\}$ to her half of the state, Bob applies $U_B \sigma U_B^{-1}$ in the former case, and $\sigma$ in the latter case, to induce decoherence in the proper stabilizer eigenstate basis.

Note that, for this protocol to work, the classical communication must be in the proper direction -- it must be Bob, not Alice, who chooses from two alternative operations. This statement sounds surprising at first, as we know that a unitary transformation applied by Bob to any maximally entangled state is equivalent to a suitable unitary transformation applied by Alice. However, unitary transformations applied by Alice and by Bob do {\em not} have equivalent effects when applied to {\em all} the elements of a maximally entangled {\em basis}.  Correspondingly, it must be Bob, not Alice, who applies the rotation to transform the stabilizer generators of eq.~(\ref{bell_stab}) to those of eq.~(\ref{twisted_bell_stab}).

\subsection{A twisted Bell basis}

We saw that the general causal complete measurement superoperator projects onto a basis that partitions ${\cal H}_A\otimes {\cal H}_B$ into $d\times d$ subspaces. Of course, if $d=1$, the basis is a product basis and the superoperator is trivially localizable. What about the other limiting case, in which $d=N_A=N_B$, so that the basis is maximally entangled? We will give an example of a $4\times 4$ maximally entangled basis, such that the corresponding measurement superoperator is {\em not} localizable.

In the $d\times d$ case,  any maximally entangled state can be expressed as $U\otimes I|\Phi^+\rangle$, where
\begin{equation}
|\Phi^+\rangle = {1\over\sqrt{d}}\sum_{i=1}^d |i\rangle\otimes |i\rangle
\end{equation} and $U$ is unitary. The elements of a $d\times d$ maximally entangled basis, then, can be expressed as
\begin{equation}
|\Phi_a\rangle = U_a\otimes I |\Phi^+\rangle~;
\end{equation}
the requirement that the states are orthogonal becomes
\begin{equation}
\label{unitary_on}
{\rm tr}\left(U_a^\dagger U_b\right) = d\,\delta_{ab}~.
\end{equation}
For the standard $d\times d$ Bell basis, these unitary transformations can be chosen
as
\begin{equation}
U_{a,b}= X^a Z^b~,\quad a,b=0,1,\dots, d-1~,
\end{equation}
where $X$ and $Z$ are the $d$-dimensional Pauli operators that act on a basis $\{|0\rangle,|1\rangle, \dots, |d-1\rangle\}$ as
\begin{eqnarray}
X|i\rangle & = & |i+1~{\rm (mod~d)}\rangle~,\nonumber\\
Z|i\rangle &=& \omega^i|i\rangle~, \quad \omega=e^{2\pi i/d}~.
\end{eqnarray}

{\em Any} measurement superoperator that projects onto a maximally entangled basis satisfies the criterion of Theorem 3 (in both directions) and is therefore causal. But if the superoperator is localizable, Theorem 5 requires the unitary transformations $\{U_a\}$ satisfying eq.~(\ref{unitary_on}) to obey further restrictions. Note that if Alice and Bob both adopt the Schmidt basis of $|\Phi_0\rangle$ as their computational bases, then $|\Phi_0\rangle= |\Phi^+\rangle$ and $U_0=I$. Then a useful characterization of the $U_a$'s is provided by: 

\noindent{\bf Theorem 6} {\em Let ${\cal U}$ be a set of $d^2$ $d\times d$ unitary matrices satisfying eq.~(\ref{unitary_on}), and let  $\cal E_{\cal U}$ be the measurement superoperator that projects onto the orthonormal basis $\{U_a\otimes I|\Phi^+\rangle, U_a\in {\cal U}\}$. Suppose that ${\cal E}_{\cal U}$ is localizable. Then if $I$, $U$, and $V$ are all contained in ${\cal U}$, so must be $e^{i\phi}UV$, for some phase $e^{i\phi}$. }

\noindent That is, ${\cal U}$ is a projective group.

\noindent {\bf Proof}: Theorem 6 follows easily from Theorem 5. First note that
\begin{equation}
M\otimes I|\Phi^+\rangle = I\otimes M^T|\Phi^+\rangle~,
\end{equation}
where $M$ is any operator, and the transpose is taken in the computational basis. Then by hypothesis, all of
\begin{equation}
|\Phi^+\rangle~,\quad U\otimes I|\Phi^+\rangle~,\quad I\otimes V^T|\Phi^+\rangle~,
\end{equation}
are eigenstates of ${\cal E}_{\cal U}$. Theorem 5 then implies that
\begin{equation}
U\otimes V^T|\Phi^+\rangle = UV\otimes I|\Phi^+\rangle~
\end{equation}
is also an eigenstate, and hence an element of the orthonormal basis, up to a phase. This proves the theorem.

Now to exhibit a causal measurement superoperator that is not localizable, it suffices to construct unitary operators that don't satisfy the projective group property specified in Theorem 6. Consider, in the $4\times 4$ case, the 16 unitary operators
\begin{equation}
\label{mismatch}
{\cal U}=\left\{
\matrix{I, & Z, & Z^2, & Z^3,\cr
X, & XZ, & XZ^2, & XZ^3,\cr
X^2, & X^2Z, & X^2Z^2, & X^2Z^3,\cr
X^3, & X^3\tilde Z, & X^3Z^2, & X^3\tilde Z Z^2\cr}
\right\}~,
\end{equation}
where $\tilde Z$ and $Z^2$ are the diagonal $4\times 4$ matrices
\begin{eqnarray}
\tilde Z & = & {\rm diag}(1,1,-1,-1)~.\nonumber\\
Z^2 &=& {\rm diag}(1,-1, 1,-1)~.
\end{eqnarray}
We can readily check that these operators obey the orthogonality condition eq.~(\ref{unitary_on}), as $I$, $\tilde Z$, $Z^2$, and 
\begin{equation}
\tilde Z Z^2= {\rm diag}(1,-1,-1,1)
\end{equation}
are all mutually orthogonal. (They are the characters of the four unitary irreducible representations of the group $Z_2\times Z_2$.) However, due to the mismatch of the fourth row of eq.~(\ref{mismatch}) with the first three rows, ${\cal U}$ does not have the projective group property required by Theorem 6. For example, $X$ and $X^2Z$ are contained in ${\cal U}$, but their product $X^3Z$ is not proportional to any element of ${\cal U}$. Therefore ${\cal E}_{\cal U}$ is not localizable.

As with any causal complete measurement superoperator, ${\cal E}_{\cal U}$ is semilocalizable. But in contrast to the preceding example, classical communication is not sufficient -- quantum communication (or equivalently, classical communication and shared entanglement) are needed to implement the operation. To prove this we can appeal to a result of \cite{beckman}: If a superoperator can be implemented with 1-way classical communication, and has a maximally entangled state as an eigenstate, then it is localizable. Since the superoperator that projects onto the twisted Bell basis has a maximally entangled state as an eigenstate, and is {\em not} localizable, we know that it can't be done with 1-way classical communication.

Further examples of twisted Bell bases are presented and discussed in \cite{beckman}.

\section{Quantum correlations and localizability}
In this section, we will use a different method to exhibit another class of causal superoperators that are not localizable. The construction exploits fundamental limitations on the strength of  correlations among the parts of a quantum system, limitations embodied by the Cirel'son inequality. An operation that produces correlations that are too strong cannot be implemented without communication among the parts.

A related observation is that correlations arising from quantum entanglement are stronger than can be achieved with shared randomness -- this is the content of Bell's theorem. We use this idea to construct examples of superoperators that can be locally implemented with prior quantum entanglement, but cannot be locally implemented with shared randomness.

\subsection{The CHSH and Cirel'son inequalities}
Suppose that Alice receives a classical input bit $x\in \{0,1\}$ and is to produce a classical output bit $a$, while Bob receives input bit $y$ and is to produce output bit $b$. Their goal is to generate output bits that are related to the input bits according to
\begin{equation}
a\oplus b = x\wedge y~,
\end{equation}
where $\oplus$ denotes the sum modulo 2 (the XOR gate) and $\wedge$ denotes the product (the AND gate).

If Alice and Bob are unable to communicate with one another, so that Alice does not know Bob's input and Bob does not know Alice's, then they will not be able to achieve their goal for all possible values of the input bits. Let $a_0, a_1$ denote the value of Alice's output if her input is $x=0,1$ and let $b_0,b_1$ denote Bob's output if his input is $y=0,1$. They would like their output bits to satisfy
\begin{eqnarray}
\label{chsh_goal}
&&a_0\oplus b_0= 0~,\nonumber\\
&&a_0\oplus b_1=0~,\nonumber\\
&&a_1\oplus b_0=0~,\nonumber\\
&&a_1\oplus b_1=1~;
\end{eqnarray}
this is impossible, since by summing the four equations we obtain 0=1. 

If Alice and Bob always choose the output $a=b=0$, then they will achieve their goal with probability $3/4$, if all possible values of the input bits are equally probable. The {\em CHSH inequality} says that, even if Alice and Bob share a table of random numbers, no higher success probability is attainable. To make the connection with the CHSH inequality as it is usually formulated \cite{peres}, define random variables with values $\pm 1$ as
\begin{eqnarray}
&&A=(-1)^{a_0}~, \quad A' = (-1)^{a_1}~,\nonumber\\
&&B=(-1)^{b_0}~,\quad B'=(-1)^{b_1}~.
\end{eqnarray}
Then the CHSH inequality says that for any joint probability distribution governing $A,A',B,B'\in \{\pm 1\}$, the expectation values satisfy
\begin{equation}
\label{chsh_inequality}
\langle AB\rangle + \langle AB'\rangle + \langle A'B\rangle -\langle A'B'\rangle \le 2~.
\end{equation}
Furthermore, if we denote by $p_{xy}$ the probability that eq.~(\ref{chsh_goal}) is satisfied when the input bits are $(x,y)$, then
\begin{eqnarray}
&&\langle AB\rangle = 2p_{00} -1~,\nonumber\\
&&\langle AB'\rangle = 2p_{01} -1~,\nonumber\\
&&\langle A'B\rangle = 2p_{10} -1~,\nonumber\\
&&\langle A'B'\rangle = 1- 2p_{11}~,
\end{eqnarray}
so that eq.~(\ref{chsh_inequality}) becomes \cite{bcvd}
\begin{equation}
{1\over 4}\left(p_{00} + p_{01} + p_{10} + p_{11} \right)\le {3\over 4}~.
\end{equation}

If Alice and Bob share quantum entanglement, they still can't satisfy eq.~(\ref{chsh_goal}) for all inputs, but they can achieve an improved success probability compared to the case where they share only randomness.  If we suppose that $A,A',B,B'$ are all Hermitian operators with eigenvalues $\pm 1$, and that Alice's operators $A$ and $A'$ commute with Bob's operators $B$ and $B'$, then the quantum mechanical expectation values obey the Cirel'son inequality \cite{peres}
\begin{equation}
\label{cir_inequality}
\langle AB\rangle + \langle AB'\rangle + \langle A'B\rangle -\langle A'B'\rangle \le 2\sqrt{2}~;
\end{equation}
the success probability then satisfies
\begin{equation}
\label{cir_prob}
{1\over 4}\left(p_{00} + p_{01} + p_{10} + p_{11} \right)\le {1\over 2} +{1\over 2\sqrt{2}}\approx .853~.
\end{equation}
Furthermore, the inequality can be saturated if the observables $A,A',B,B'$ are chosen appropriately.

\subsection{A causal operation that is not localizable}
Our observations concerning the Cirel'son inequality quickly lead us to a construction of a causal operation that is not localizable. 

For a two-qubit state shared by Alice and Bob, consider a superoperator, denoted ${\cal E}_\wedge$, that can be implemented in two steps. The first step is a complete orthogonal measurement that projects onto the product basis $\{|00\rangle,|01\rangle,|10\rangle,|11\rangle\}$. Then in the second step the product state is transformed according to
\begin{eqnarray}
\label{magic_box}
\matrix{|00\rangle\cr |01\rangle\cr |10\rangle\cr}\Bigg\}\to&& {1\over 2}\left(|00\rangle\langle 00| + |11\rangle\langle 11| \right) ~, \nonumber\\
|11\rangle\to&& {1\over 2} \left(|01\rangle\langle 01| + |10\rangle\langle 10|\right)~.
\end{eqnarray}
This operation is obviously trace preserving, and since it has an evident operator-sum representation, it is also completely positive. Furthermore, it is causal. Whatever the initial state that Alice and Bob share, each has the final density operator $\rho=I/2$; therefore, neither can receive a signal from the other.

Though causal, ${\cal E}_\wedge$ is not localizable -- it cannot be implemented by Alice and Bob without communication, even if they share an entangled ancilla. If it were localizable, then Alice and Bob would be able to implement ${\cal E}_\wedge$ by applying a local unitary transformation $U_A\otimes U_B$ to the composite system consisting of the input qubits and ancilla, and then throwing some qubits away. Let the input state shared by Alice and Bob be one of the products states $\{|00\rangle, |01\rangle, |10\rangle, |11\rangle\}$, let them apply their local unitary transformation to implement ${\cal E}_\wedge$, and suppose that each measures her or his output qubit in the basis $\{|0\rangle,|1\rangle\}$ after ${\cal E}_\wedge$ is performed. In effect then, Alice subjects the initial state to a measurement of the observable $U_A^{-1}Z_{A, {\rm out}}U_A$, and Bob measures $U_B^{-1}Z_{B, {\rm out}}U_B$, where $Z_{\cdot, {\rm out}}$ denotes a Pauli operator acting on an output qubit. Both observables have eigenvalues $\pm 1$.

Now, the Cirel'son inequality applies to a situation where Alice measures either of two observables in a specified state, and Bob does likewise. Here we are considering a case in which Alice and Bob measure fixed observables, and the initial state can be any of four possible states. But either scheme can be easily related to the other. For example, instead of providing Alice with an input qubit that can be $|0\rangle$ or $|1\rangle$, we can give her the input $|0\rangle$ and instruct her to apply $X$, or not, before she performs her measurement. In this scenario, Alice receives a classical input bit that instructs her to measure one of the two observables
\begin{eqnarray}
A&=& U_A^{-1}Z_{A, {\rm out}}U_A~,\nonumber\\
A'&=& X_{A, {\rm in}}U_A^{-1}Z_{A, {\rm out}}U_AX_{A, {\rm in}}~,
\end{eqnarray}
and similarly for Bob. 

In this case, then, the Cirel'son inequality constrains how Alice's measurement outcome $|a\rangle$ is correlated with Bob's measurement outcome $|b\rangle$. But if they have really succeeded in implementing the operation ${\cal E}_\wedge$, then the outcomes are related to the classical input bits $x,y$ by $a\oplus b=x\wedge y$ with probability 1, a violation of the bound eq.~(\ref{cir_prob}). We conclude that no local protocol implementing ${\cal E}_\wedge$ is possible.

However, it is also clear that ${\cal E}_\wedge$ is {\em semilocalizable} -- it can be implemented with one-way {\em classical} communication from Alice to Bob (or from Bob to Alice). Alice measures her qubit in the basis $\{|0\rangle,|1\rangle\}$, and she tosses a coin to decide whether to flip her qubit or not. Then she sends her measurement result and the outcome of her coin toss to Bob. Bob measures his qubit in the basis $\{|0\rangle,|1\rangle\}$, and after reading the data sent by Alice, either flips it or not. Bob arranges that his qubit have the same value as Alice's unless they both measure $|1\rangle$, in which case he arranges for his qubit and Alice's to have opposite values. This procedure implements ${\cal E}_\wedge$.

How much communication is necessary? As pointed out to us by Cleve \cite{cleve_private}, we can obtain a lower bound on the amount of communication needed to implement ${\cal E}_\wedge$ from known lower bounds on the quantum communication complexity of the inner product function \cite{inner_product}.

Suppose that Alice has an $n$-bit classical input string $x=(x_1,x_2, \dots, x_n)$, not known to Bob, and Bob has an $n$-bit classical input string $y=(y_1, y_2, \dots, y_n)$, not known to Alice. Their goal is to compute the mod 2 inner product of their strings,
\begin{equation}
{\rm IP}(x,y)= x_1y_1 \oplus x_2y_2 \oplus \cdots \oplus x_ny_n~.
\end{equation}
It is known \cite{inner_product} that even if Alice and Bob share pre-existing entanglement, neither party can evaluate ${\rm IP}(x,y)$ with zero probability of error unless at least $n/2$ qubits are transmitted between the parties. For $n$ even, $n/2$ qubits of communication are also sufficient: Alice can use superdense coding to send $x$ to Bob, and Bob can then evaluate ${\rm IP}(x,y)$.

But if Alice and Bob were able to implement ${\cal E}_\wedge$ ``for free,'' they could use it to evaluate ${\rm IP}(x,y)$ at a smaller communication cost. Alice prepares the $n$-qubit state $|x\rangle$ and Bob the $n$-qubit state $|y\rangle$. Then ${\cal E}_\wedge$ is applied to $|x_i,y_i\rangle$ for each $i=1,2,\dots, n$, and Alice and Bob measure their qubits to obtain outputs $a_i, b_i$ for each $i$. Since $a_i\oplus b_i = x_i y_i$, we see that
\begin{eqnarray}
&&{\rm IP}(x,y)= (a_1\oplus b_1)\oplus (a_2\oplus b_2) \oplus \cdots \oplus (a_n\oplus b_n)~\nonumber\\
&&= (a_1\oplus a_2\oplus \cdots \oplus a_n)
\oplus (b_1\oplus b_2 \oplus \cdots \oplus b_n)~.
\end{eqnarray}
Therefore, Alice can evaluate the sum (mod 2) of her $n$ measurement outcomes, and send the {\em one-bit} result to Bob. Bob adds Alice's result to the sum of his own measurement outcomes, and so obtains the value of ${\rm IP}(x,y)$. Just one bit of communication is required.

Suppose that Alice and Bob have a protocol that allows them to implement ${\cal E}_\wedge$ with, on average, $Q_{\rm av}$ qubits of quantum communication. (Alice's decision whether to send a qubit could be conditioned on the outcome of a local measurement; therefore the amount of communication required can fluctuate about this average.)  Now, if Alice and Bob can implement ${\cal E}_\wedge$ $n$ times with $Q_n$ qubits of communication, then since just one additional bit is needed to complete the evaluation of the inner product function, we know that 
\begin{equation}
Q_n + 1 \ge n/2~.
\end{equation}
For large $n$, $Q_n$ converges to $nQ_{\rm av}$, and we conclude that 
\begin{equation}
Q_{\rm av}\ge 1/2~.
\end{equation}
 
This argument illustrates a general approach to proving that a quantum operation is unlocalizable: if implementing the operation would allow us to reduce the communication complexity of a function below established lower bounds, then no local implementation is possible.

\subsection{Entanglement is stronger than shared randomness}

The separation between the CHSH and Cirel'son inequalities allows us to construct a class of operations that can be implemented locally with shared entanglement, but cannot be implemented locally with shared randomness. (The existence of operations with this property was already pointed out in \S IIC.)

Suppose that Alice and Bob have a shared maximally entangled pair of qubits (qubits 3 and 4), as well as two input qubits (1 and 2) on which the operation is to act. Alice measures qubit 1 and Bob measures qubit 2 in the basis $\{|0\rangle, |1\rangle\}$. Then Alice measures her half of the entangled pair, qubit 3, choosing to measure the observable $A$ if the measured input was $|0\rangle_1$ or $A'$ if the measured input was $|1\rangle_1$. Similarly, Bob measures qubit 4, choosing to measure either $B$ or $B'$ depending on the outcome of his measurement of the input qubit 2. 

After measuring $A$ (or $A'$), Alice rotates qubit 3 to the state $|0\rangle_3$ if she found $A=1$ (or $A'=1$), and rotates it to $|1\rangle_3$ if she found $A=-1$ (or $A'=-1$). Bob does the same to qubit 4. Finally Alice and Bob throw away the input qubits 1 and 2, retaining qubits 3 and 4.

Alice and Bob, then, using their shared entanglement, have locally implemented an operation that acts on a product-state input and produces a product-state output, according to 
\begin{equation}
|x,y\rangle \to |a,b\rangle~.
\end{equation}
Averaged over the four possible product state inputs, the output of the operation satisfies $a\oplus b=xy$ with a success probability that we'll call $p$. If the observables $A$, $A'$, $B$, $B'$ are chosen to saturate the Cirel'son inequality, then $p=\cos^2(\pi/8)\approx .853$.

As is well known \cite{peres}, probability distributions for quantum
measurements of a single qubit can be correctly accounted for
by a ``hidden variable theory'' (while measurements of entangled
qubits cannot be). Therefore, measurements performed by Alice and Bob on a product input state can be perfectly simulated by a classical probability distribution, so that the measurement results must respect the CHSH inequality, which requires that the success probability $p$ satisfy $p\le 3/4$. For $3/4 < p \le \cos^2(\pi/8)$, the operation can be implemented locally with shared entanglement, but not with shared randomness.

\section{Unitarity and causality}

An important special case of an operation is a unitary transformation. In this case, our classification collapses -- the classes of causal, localizable, semicausal, and semilocalizable unitary transformations all coincide, according to

\noindent{\bf Theorem 7} {\em A bipartite unitary transformation $U_{AB}$ is semicausal if and only if it is a tensor product, $U_{AB}=U_A\otimes U_B$.}

\noindent {\bf Proof}: It's obvious that a tensor product transformation is causal.  The nontrivial content of the theorem is that if $U_{AB}$ is not a tensor product, then we can devise protocols whereby Alice can signal Bob and Bob can signal Alice; hence $U_{AB}$ is not semicausal.

To prove this, we first recall that since linear operators are a vector space with a Hilbert-Schmidt inner product, a bipartite operator (whether unitary or not) can be Schmidt decomposed \cite{nielsen_phd}.
We may write
\begin{eqnarray}
  U_{AB} = \sum_\mu \lambda_\mu A_\mu \otimes B_\mu~,
\end{eqnarray}
where the $\lambda_\mu$'s are nonnegative real numbers, and the operator bases $\{A_\mu\}$ and $\{B_\mu\}$ are orthogonal:
\begin{eqnarray} \label{eq:orthog}
  {\rm tr}(A_\mu^{\dagger} A_\nu) = N_A\delta_{\mu\nu} ~, \quad {\rm tr}(B_\mu^{\dagger}
  B_\nu) = N_B\delta_{\mu\nu} ~;
\end{eqnarray}
$N_A$ is the dimension of ${\cal H_A}$ and $N_B$ is the dimension of ${\cal H}_B$ -- we've chosen this normalization so that unitary operators are properly normalized.

If $U_{AB}$ is not a tensor product, than more than one $\lambda_\mu$ is strictly positive.  
We will show that if this is true, then $U_{AB}$ allows Bob to signal Alice. (A similar argument shows that Alice can signal Bob.) Suppose that Alice introduces a {\em reference system} ${\cal H}_R$ and that she prepares a maximally entangled state of ${\cal H}_R\otimes {\cal H}_A$
\begin{equation}
|\Phi\rangle_{RA}=\sum_i |i\rangle_R\otimes |i\rangle_A~.
\end{equation}
(Because it will be convenient later on, we have chosen an unconventional normalization of the state $|\Phi\rangle_{RA}$.)
Meanwhile, Bob prepares a pure state $|\psi\rangle_B$. When $U_{AB}$ acts, the density operator becomes 
\begin{eqnarray}
 \rho_{RAB}= \sum_{\mu\nu} \lambda_\mu \lambda_\nu ~(I\otimes A_\mu)|\Phi\rangle \langle \Phi| (I\otimes A_\nu^{\dagger}) \nonumber\\
\otimes \,B_\mu |\psi\rangle \langle \psi| B_\nu^{\dagger}
  ~.
\end{eqnarray}
After tracing out Bob's system, the density operator of Alice's system becomes
\begin{eqnarray} \label{eq:unitary_crucial}
  \rho_{RA} =
  \sum_{\mu\nu} \lambda_\mu \lambda_\nu (I\otimes A_\mu)|\Phi\rangle \langle \Phi| (I\otimes A_\nu^{\dagger})\nonumber\\
\cdot\, \langle \psi| B_\nu^{\dagger} B_\mu | \psi
  \rangle \,\, ~.
\end{eqnarray}

Bob can signal Alice if the density operator $\rho_{RA}$ depends on Bob's initial state $|\psi\rangle$. It follows from eq.~(\ref{eq:orthog}) that the states $\{(I\otimes A_\mu) |\Phi\rangle\}$ are mutually orthogonal; therefore signaling is possible if there exist two states $|\psi\rangle$ and $|\psi'\rangle$ such that
\begin{eqnarray}
\label{munune}
    \langle \psi| B_\nu^{\dagger} B_\mu |\psi\ra \neq \la \psi'| B_\nu^{\dagger}
    B_\mu |\psi'\rangle~.
  \end{eqnarray}
for some $\mu$ and $\nu$.

Now we distinguish two cases:
\begin{enumerate}
\item Suppose that for some $\mu$, $B_\mu$ is not unitary. Then, in order to satisfy the normalization condition eq.~(\ref{eq:orthog}), $B_\mu^\dagger B_\mu$ must have (at least) two distinct eigenvalues. Choose $|\psi\rangle$ and $|\psi'\rangle$ to be the corresponding eigenvectors. Then eq.~(\ref{munune}) is satisfied for $\nu=\mu$.
  
\item Suppose that $B_\mu$ and $B_\nu$ are both unitary for $\mu\ne\nu$. Then $B_\nu^\dagger B_\mu$ is nonzero and (according to eq.~(\ref{eq:orthog})) has vanishing trace; therefore it has (at least) two distinct eigenvalues. Thus eq.~(\ref{munune}) is satisfied, where $|\psi\rangle$ and $|\psi'\rangle$ are the corresponding eigenvectors. 
\end{enumerate}

\noindent In either case, Alice's density operator depends on how Bob's initial state is chosen; hence Bob can signal Alice. A similar argument shows that Alice can signal Bob. Therefore $U_{AB}$ is not semicausal. This completes the proof of Theorem 7.

It follows immediately from Theorem 7 that if a bipartite unitary transformation is semicausal, it is also localizable, and therefore fully causal.

\section{General operations: criteria for semicausality}

To show that an operation ${\cal E}$ is {\em not} fully causal, it suffices to exhibit a protocol whereby the operation can be used to send a signal in one direction, and to show that it is not semicausal, it suffices to exhibit protocols for signaling in both directions. On the other hand, to show that it {\em is} fully causal (or semicausal), we must prove that no such signaling protocols exist. To settle whether a particular operation is causal, it is very helpful to have a simpler criterion that can be checked with a straightforward calculation. We will now develop such a criterion for semicausal superoperators.

First, we recall that, although in our definition of semicausality we allowed the initial state shared by Alice and Bob to be entangled, we could without loss of generality restrict their initial state to be a product state (Theorem 1). Next, we note that a helpful tool in our analysis of the causality properties of unitary transformations, entanglement with a reference system, can also be fruitfully applied to the general case. To give a useful restatement of the criterion for semicausality, suppose again that Alice prepares the maximally entangled state $|\Phi\rangle_{RA}$ of her reference system ${\cal H}_R$ with her system ${\cal H}_A$. Then for each state $|\varphi\rangle_A\in {\cal H}_A$, there is a corresponding ``relative state''  $|\varphi^*\rangle_R\in {\cal H}_R$, chosen so that 
\begin{equation}
{}_R\langle \varphi^*|\Phi\rangle_{RA}= |\varphi\rangle_A~.
\end{equation}
We can easily see that an operation ${\cal E}$ is semicausal (Bob is unable to signal Alice) if and only if
\begin{eqnarray}
\label{crit_alice_ref}
{\rm tr}_B\big({\cal E}(|\Phi\rangle\langle \Phi|\otimes |\psi\rangle\langle\psi|)\big)
\end{eqnarray}
is independent of Bob's state $|\psi\rangle$. (Here of course ${\cal E}$ really denotes the operation $I_R\otimes {\cal E}_{AB}$ acting on the $RAB$ system.) If the expression in eq.~(\ref{crit_alice_ref}) depends on $|\psi\rangle$, then obviously Bob can signal Alice.  Conversely, if Bob can signal Alice, then there is a signaling protocol in which the initial state is a product of pure states; there are states  $|\varphi\rangle$, $|\psi\rangle$, and  $|\psi'\rangle$, such that 
\begin{eqnarray}
&&{}_R\langle\varphi^*|{\rm tr}_B\big({\cal E}(|\Phi\rangle\langle \Phi|\otimes |\psi\rangle\langle\psi|)\big)|\varphi^*\rangle_R\nonumber\\
&&={\rm tr}_B\big({\cal E}(|\varphi\rangle \langle \varphi|\otimes |\psi\rangle\langle\psi|)\big)\nonumber\\
&&\ne {\rm tr}_B\big({\cal E}(|\varphi\rangle \langle \varphi|\otimes |\psi'\rangle\langle\psi'|)\big)\nonumber\\
&&={}_R\langle\varphi^*|{\rm tr}_B\big({\cal E}(|\Phi\rangle\langle \Phi|\otimes 
|\psi'\rangle\langle\psi'|)\big)|\varphi^*\rangle_R~.
\end{eqnarray}
Since we have found a particular matrix element of ${\rm tr}_B\big({\cal E}(|\Phi\rangle\langle \Phi|\otimes |\psi\rangle\langle\psi|)\big)$ that depends on $|\psi\rangle$, evidently so does ${\rm tr}_B\big({\cal E}(|\Phi\rangle\langle \Phi|\otimes |\psi\rangle\langle\psi|)\big)$ itself.

Now, let's provide Bob with a reference system $S$, and suppose that he prepares a maximally entangled state of ${\cal H}_B\otimes {\cal H}_S$
\begin{equation}
|\Phi'\rangle_{BS}= \sum_i |i\rangle_B\otimes |i\rangle_S~
\end{equation}
(again we have chosen an unconventional normalization for convenience). We are ready to state and prove our new criterion for semicausality:

\noindent{\bf Theorem 8} {\em Let $|\Phi\rangle_{RA}$ be a maximally entangled state of system $A$ with the reference system $R$, and let $|\Phi'\rangle_{BS}$ be a maximally entangled state of system $B$ with reference system $S$. Then the bipartite superoperator ${\cal E}$ acting on $AB$ is semicausal (Bob cannot signal Alice) if and only if
\begin{equation}
{\rm tr}_B\bigg(({\cal E}_{AB}\otimes I_{RS}) \Big(\left(|\Phi\rangle\langle\Phi|\right)_{RA}\otimes \left(|\Phi'\rangle\langle\Phi'|\right)_{BS}\Big)\bigg)
\end{equation}
is proportional to the product $\rho_{RA}\otimes I_S$~, where $I_S$ denotes the identity on $S$.}

\noindent {\bf Proof:}
If ${\rm tr}_B({\cal E}(|\Phi\rangle\langle\Phi|\otimes |\Phi'\rangle\langle\Phi'|))$ is proportional to
$\rho_{RA}\otimes I_S$, then by evaluating the matrix element between relative states $|\psi^*\rangle_S$, we see that ${\rm tr}_B\big({\cal E}(|\Phi\rangle\langle \Phi|\otimes |\psi\rangle\langle\psi|)\big)$ is independent of $|\psi\rangle$. Therefore Bob cannot signal Alice. Conversely, suppose that Bob cannot signal Alice. Then 
\begin{eqnarray}
&&{\rm tr}_B\big({\cal E}(|\Phi\rangle\langle \Phi|\otimes |\psi\rangle\langle\psi|)\big)\nonumber\\
&=& {}_S\langle\psi^*| {\rm tr}_B\Big({\cal E}\big(|\Phi\rangle\langle\Phi|\otimes |\Phi'\rangle\langle\Phi'|\big)\Big)|\psi^*\rangle_S
\end{eqnarray}
is independent of $|\psi^*\rangle$. It follows that ${\rm tr}_B({\cal E}(|\Phi\rangle\langle\Phi|\otimes |\Phi'\rangle\langle\Phi'|))$ is proportional to $I_S$. This proves Theorem 8.

\section{Conclusions}

We have studied the constraints on quantum operations that are imposed by relativistic causality. In the bipartite setting where no classical communication is permitted, we find a hierarchy of operations:
\begin{description}
\item[(1)] operations that can be implemented with no shared resources,
\item[(2)] operations that can be implemented with shared randomness,
\item[(3)] operations that can be implemented with shared entanglement (localizable operations),
\item[(4)] causal operations,
\item[(5)] acausal operations.
\end{description}
Our central observation is that the classes (3) and (4) do not coincide: there are operations that respect causality, but are nonetheless forbidden by the rules of local quantum physics. 

Our work can be regarded as a useful step toward the broader goal of characterizing the physically realizable operations in relativistic quantum field theory. However, as noted in the Sec. I, to apply our results to field theory one must accept the idealization that the resources shared by the parties are external probes not themselves described by the field theory. 

In a separate paper, we have also discussed causality constraints that apply to non-Abelian gauge theories \cite{preskill}; we have shown that the {\em nondemolition} measurement of a spacelike Wilson loop operator is an acausal operation (confirming a speculation of Sorkin \cite{sorkin}), and is therefore surely not localizable. On the other hand, a {\em destructive} measurement of a Wilson loop is possible --- spacelike separated parties can perform a POVM from which the value of the Wilson loop can be inferred, but this POVM will damage Wilson loop eigenstates. 

The compatibility of quantum mechanics with special relativity is highly nontrivial; in fact, it is something of a miracle. Because relativistic quantum field theories are so highly constrained, it is tempting to speculate that ``quantum mechanics is the way it is because any small changes in quantum mechanics would lead to absurdities'' \cite{weinberg}.

From this perspective, the existence of causal operations that are not localizable comes as a surprise. We seem to have the freedom to relax the rules of quantum theory by allowing more general operations, without encountering unacceptable physical consequences. Nontrivial support for this notion is provided by the semigroup property of the causal operations. It is reasonable to insist that the operations allowed at a given time ought not to depend on the previous history of the system; since the composition of two causal operations is causal, a theory that admits more general causal operations than those allowed in local quantum theory could adhere to this proviso.

One wonders whether there are further principles, beyond relativistic causality, that will restrict the class of allowed operations to those and only those that are truly realizable in Nature. If so, these principles might lead us to an understanding of why quantum mechanics has to be the way it is. What might these principles be?

We don't know. But the discussion in \S VI invites us to contemplate the fundamental limitations on the correlations among the parts of a physical system. Experimentally confirmed violations of the CHSH inequality demonstrate that the correlations are stronger than those allowed by any local hidden variable theory. Operations that are causal but not localizable produce correlations that are stronger still, and violate the Cirel'son inequality. What criteria point toward a description of Nature that incorporates violation of the CHSH inequality, but not violation of the Cirel'son inequality?

Or could it be that Nature really does allow more general operations, and that the conventional framework of local quantum physics needs revision? Ultimately, only experiment can decide.

Note added: After this paper appeared, a proof of the conjecture that semicausal superoperators are semilocalizable was found by Eggeling, Schlingemann, and Werner \cite{eggeling}.
\acknowledgments

We thank Harry Buhrman, Richard Cleve, and David DiVincenzo for helpful discussions, Jennifer Dodd for comments on the manuscript, and Reinhard Werner and Rainer Verch for instructive correspondence.
This work has been supported in part by the Department of Energy under Grant No. DE-FG03-92-ER40701, by the National Science Foundation, and by an IBM Faculty Partnership Award. Some of this work was done at the Aspen Center for Physics. This research was partially conducted during the period that D.G. served as a Clay Long-Term CMI Prize Fellow. 

\appendix
\section{Proof of Theorem 3}

\noindent {\bf Theorem 3} {\em Consider a bipartite complete orthogonal measurement superoperator of the form
\begin{equation}
{\cal E}(\rho) = \sum_a |a\rangle\langle a|\rho|a\rangle\langle a|~,
\end{equation}
where $\{|a\rangle\}$ is an orthonormal basis for ${\cal H}_A\otimes {\cal H}_B$, and let $\sigma_A^a= {\rm tr}_B(|a\rangle\langle a|)$. Then ${\cal E}$ is semicausal if and only if the following property holds: For each pair of operators $\{\sigma_A^a,\sigma_A^b\}$, either $\sigma_A^a=\sigma_A^b$ or $\sigma_A^a\sigma_A^b=0$.}

\noindent{\bf Proof}: We begin by observing that if $\sigma_A^a\sigma_A^b=0$, then
\begin{equation}
\label{unitary_orthogonal}
\langle a|I_A\otimes U_B|b\rangle=0
\end{equation}
for any unitary $U_B$. To show this, we Schmidt decompose the states $|a\rangle$ and $|b\rangle$:
\begin{eqnarray}
\label{schmidt_decomp}
|a\rangle &=& \sum_i \sqrt{\lambda_{a,i}} |a,i\rangle_A\otimes |a,i\rangle_B~,\nonumber\\
|b\rangle &=& \sum_i \sqrt{\lambda_{b,i}} |b,i\rangle_A\otimes |b,i\rangle_B~,
\end{eqnarray}
where $\{|a,i\rangle_A\}$, $\{|b,i\rangle_A\}$ are orthonormal bases of ${\cal H}_A$, $\{|a,i\rangle_B\}$, $\{|b,i\rangle_B\}$ are orthonormal bases of ${\cal H}_B$, and the $\lambda_{a,i}$'s, $\lambda_{b,i}$'s are all nonnegative. In terms of these bases, we find
\begin{eqnarray}
\sigma_A^a&=&\sum_i \lambda_{a,i}|a,i\rangle_A {}_A\langle a,i|~,\nonumber\\
\sigma_A^b&=&\sum_i \lambda_{b,i}|b,i\rangle_A {}_A\langle b,i|~;
\end{eqnarray}
therefore $\sigma_A^a\sigma_A^b=0$ iff ${}_A\langle a,i|b,j\rangle_A=0$ for each $i$ and $j$ with $\lambda_{a,i}\lambda_{b,j}\ne 0$. Eq.~(\ref{unitary_orthogonal}) follows immediately.

Now suppose that for each $a$ and $b$, either $\sigma_A^a=\sigma_A^b$ or $\sigma_A^a\sigma_A^b=0$. Let $|\psi\rangle$ be an arbitrary pure state in ${\cal H}_A\otimes {\cal H}_B$. We will show that 
\begin{eqnarray}
\label{unitary_1causal}
&&{\rm tr}_B~{\cal E}(|\psi\rangle\langle \psi|)\nonumber\\
&&\quad = {\rm tr}_B~{\cal E}\big((I_A\otimes U_B)|\psi\rangle\langle \psi|(I_A\otimes U_B^\dagger)\big)
\end{eqnarray}
for any unitary $U_B$, which according to Theorem 1 suffices to show that ${\cal E}$ is semicausal.
To prove eq.~(\ref{unitary_1causal}), we expand the state $|\psi\rangle$ in the basis $\{|b\rangle\}$ as
\begin{equation}
|\psi\rangle = \sum_b \alpha_b|b\rangle~,
\end{equation}
and so obtain 
\begin{eqnarray}
\label{unitary_sum}
&&{\rm tr}_B~{\cal E}\big((I_A\otimes U_B)|\psi\rangle\langle \psi|(I_A\otimes U_B^\dagger)\big)\nonumber\\
&& = \sum_{a,b,c} \alpha_b\alpha_c^* \langle a|I_A\otimes U_B|b\rangle\langle c|I_A\otimes U_B^\dagger|a\rangle \cdot \sigma_A^a~.
\end{eqnarray}
Now we use the property that either $\sigma_A^a=\sigma_A^b$ or $\sigma_A^a\sigma_A^b=0$ for each $a$ and $b$: because $\sigma_A^a\sigma_A^b=0$ implies that $\langle a|I_A\otimes U_B|b\rangle=0$, we can replace $\sigma_A^a$ by $\sigma_A^b$ in eq.~(\ref{unitary_sum}) without altering the sum. After this replacement, we use the property $\sum_a|a\rangle\langle a|=I$ and the unitarity of $U_B$ to find
\begin{eqnarray}
&&{\rm tr}_B~{\cal E}\big((I_A\otimes U_B)|\psi\rangle\langle \psi|(I_A\otimes U_B^\dagger)\big)\nonumber\\
&& = \sum_{b} |\alpha_b|^2\cdot \sigma_A^b~, 
\end{eqnarray}
which is independent of $U_B$, proving eq.~(\ref{unitary_1causal}) and hence the ``if'' part of Theorem 3.

To prove the ``only if'' part of Theorem 3, we suppose that for some $a$ and $b$, $\sigma_A^a\sigma_A^b\ne 0$ and $\sigma_A^a\ne \sigma_A^b$; we must show that Bob can signal Alice. It suffices to show that a basis element $|b\rangle$ and a unitary transformation $U_B$ can be chosen so that
\begin{eqnarray}
\label{signal_exists}
&&{\rm tr}_B{\cal E}\big((I_A\otimes U_B)|b\rangle\langle b|(I_A\otimes U_B^\dagger)\big)\nonumber\\
&&\quad =\sum_a |\langle a|I_A\otimes U_B|b\rangle|^2 \cdot \sigma_A^a\nonumber\\
&&\quad \ne \sigma_A^b ={\rm tr}_B{\cal E}(|b\rangle\langle b|)~.
\end{eqnarray}
If eq.~(\ref{signal_exists}) holds, then Bob can signal Alice by the following protocol: Alice and Bob prepare in advance the shared state $|b\rangle$. Just before ${\cal E}$ acts, Bob either applies $U_B$ to the state or he does nothing. Eq.~(\ref{signal_exists}) says that Alice's density operator after ${\cal E}$ acts depends on the action chosen by Bob; therefore, Bob can signal Alice.

We will prove in two steps that $U_B$ and $|b\rangle$ exist such that eq.~(\ref{signal_exists}) is satisfied. The first step is to show that for $\sigma_A^a\sigma_A^b\ne 0$, there is a unitary $U_B$ such that
\begin{equation}
\langle b| I_A\otimes U_B |a\rangle\ne 0~.
\end{equation}
In terms of the Schmidt bases defined in eq.~(\ref{schmidt_decomp}), what is to be shown can be rewritten as 
\begin{equation}
\sum_{i,j}\sqrt{\lambda_{b,j}\lambda_{a,i}} ~{}_A\langle b,j|a,i\rangle_A\cdot {}_B\langle b,j|U_B|a,i\rangle_B\ne 0~.
\end{equation}
Now recall that $\sigma_A^a\sigma_A^b\ne 0$ implies that ${}_A\langle b,j|a,i\rangle_A\ne 0$ for some $i$ and $j$. By labeling the Schmidt bases appropriately we can ensure that ${}_A\langle b,1|a,1\rangle_A\ne 0$. By adopting suitable phase conventions, we can ensure that each ${}_A\langle b,i|a,i\rangle_A$ is real and nonnegative, and we can choose $U_B$ so that ${}_B\langle b,j|U_B|a,i\rangle_B=\delta_{ij}$. Thus 
\begin{eqnarray}
&&\sum_{i,j}\sqrt{\lambda_{b,j}\lambda_{a,i}} ~{}_A\langle b,j|a,i\rangle_A\cdot {}_B\langle b,j|U_B|a,i\rangle_B \nonumber\\
&&\quad =  \sum_{i}\sqrt{\lambda_{b,i}\lambda_{a,i}}~{}_A\langle b,i|a,i\rangle_A
\end{eqnarray}
is a sum of nonnegative terms, at least one of which is nonzero; therefore the sum is surely nonzero, as we wished to show.

Now we have seen that $U_B$ and $|b\rangle$ can be chosen so that the sum in eq.~(\ref{signal_exists}) contains a term other than $\sigma_A^b$. The second step of the argument will establish that we can, in fact, choose $U_B$ and $|b\rangle$ such that the sum is not equal to $\sigma_A^b$.

For this purpose, consider the set $S^b$ containing all $\sigma_A^a$ such that $\sigma_A^a\sigma_A^b\ne 0$. Suppose that $S^b$ contains at least two elements, and that $\sigma_A^b$ is an {\em extremal} element of $S^b$ -- that is $\sigma_A^b$ cannot be expressed as a nontrivial convex combination of other elements of $S^b$. Then since the sum in eq.~(\ref{signal_exists}) is a convex combination of elements of $S^b$, and since we can choose $U_B$ so that the sum contains some $\sigma_A^a\ne \sigma_A^b$ with a nonvanishing coefficient, the inequality in eq.~(\ref{signal_exists}) follows from the extremality of $\sigma_A^b$ in $S^b$.  

Finally, it only remains to show that $|b\rangle$ can be chosen so that $\sigma_A^b$ is extremal in $S^b$. For this purpose, of all $\sigma_A^b$ such that $S^b$ contains two or more elements, choose one with maximal Hilbert-Schmidt norm ({\it i.e.}, with maximal  ${\rm tr}\left[(\sigma_A^b)^2\right]$). We claim that this $\sigma_A^b$ must be extremal in $S^b$. 

To see that $\sigma_A^b$ is extremal in $S^b$, we appeal to the following property: Let $\{v_i\}$ be a finite set of vectors, and let $\parallel v \parallel_{\rm max}$ be the maximum value of $\parallel v_i\parallel$. Then the {\em strict} inequality 
\begin{equation}
\label{convex_norm}
\parallel \sum_{i} p_i v_i\parallel ~<~ \parallel v\parallel_{\rm max}
\end{equation}
holds for any nontrivial convex combination of the $v_i$'s (one with two or more nonvanishing $p_i$'s). Applying eq.~(\ref{convex_norm}) to $S^b$, the Hilbert-Schmidt norm of our
selected $\sigma_A^b$ is on the right-hand side, which is
strictly greater than the left-hand side, the norm of any nontrivial 
convex combination of elements of $S^b$.  Therefore $\sigma_A^b$
is extremal in $S^b$.

This completes the proof of Theorem 3.
\section{Proof of Theorem 5}

\noindent {\bf Theorem 5} {\em If ${\cal E}$ is a localizable superoperator on ${\cal H}_A\otimes {\cal H}_B$, and $|\psi\rangle$, $A\otimes I|\psi\rangle$, and $I\otimes B|\psi\rangle$ are all eigenstates of ${\cal E}$ (where $A$ and $B$ are invertible operators), then $A\otimes B|\psi\rangle$ is also an eigenstate of ${\cal E}.$}

\noindent {\bf Proof:} If the superoperator ${\cal E}$ is localizable, its action on a pure state $|\eta\rangle_{AB}$ can be realized by a tensor product unitary transformation $U_{RA}\otimes V_{BS}$ acting on $|\eta\rangle_{AB}|\varphi\rangle_{RS}$, where $|\varphi\rangle_{RS}$ is a suitable ancilla state shared by Alice and Bob. By hypothesis, this unitary transformation acting on $|\psi\rangle_{AB}|\varphi\rangle_{RS}$ preserves $|\psi\rangle_{AB}$ and rotates only the ancilla state:
\begin{equation}
U_{RA}\otimes V_{BS}|\psi\rangle_{AB}|\varphi\rangle_{RS}= |\psi\rangle_{AB}|\varphi_0\rangle_{RS}~,
\end{equation}
for some state of the ancilla $|\varphi_0\rangle_{RS}$. Similarly, by hypothesis, we have
\begin{eqnarray}
&&U_{RA}\otimes V_{BS}(A_A\otimes I_B)|\psi\rangle_{AB}|\varphi\rangle_{RS}\nonumber\\
&&\qquad = (A_A\otimes I_B)|\psi\rangle_{AB}|\varphi_A\rangle_{RS}~,\nonumber\\
&&U_{RA}\otimes V_{BS}(I_A\otimes B_B)|\psi\rangle_{AB}|\varphi\rangle_{RS}\nonumber\\
&&\qquad = (I_A\otimes B_B)|\psi\rangle_{AB}|\varphi_B\rangle_{RS}~,
\end{eqnarray}
for states of the ancilla $|\varphi_A\rangle_{RS}$, $|\varphi_B\rangle_{RS}$.

Now consider the transformation $\Delta_{RA}= UAU^{-1}A^{-1}$. By construction, $\Delta_{RA}$ acts only on Alice's system $RA$. In fact, we can show that when acting on the state $(A_A\otimes I_B)|\psi\rangle_{AB}|\varphi_0\rangle_{RS}$, $\Delta_{RA}$ acts trivially on $A$ and nontrivially only on Alice's ancilla $R$; we observe that
\begin{eqnarray}
&&(A_A\otimes I_B)|\psi\rangle_{AB}|\varphi_A\rangle_{RS} \nonumber\\
&&\qquad = (U_{RA}\otimes V_{BS})(A_A\otimes I_B)|\psi\rangle_{AB}|\varphi\rangle_{RS}\nonumber\\
&&\qquad = (\Delta_{RA}\otimes I_B)(A_A\otimes I_B)(U_{RA}\otimes V_{BS})|\psi\rangle_{AB}|\varphi\rangle_{RS}\nonumber\\
&&\qquad = (\Delta_{RA}\otimes I_B)(A_A\otimes I_B)|\psi\rangle_{AB}|\varphi_0\rangle_{RS}~.
\end{eqnarray}
Therefore, acting on $(A_A\otimes I_B)|\psi\rangle_{AB}|\varphi_0\rangle_{RS}$, we may replace $\Delta_{RA}\otimes I_B$ by $R_R\otimes I_B$, where $R_R$ is a (unitary) transformation acting on $R$ alone that rotates $|\varphi_0\rangle_{RS}$ to $|\varphi_A\rangle_{RS}$. We then have
\begin{eqnarray}
&&(U_{RA}A_A\otimes V_{BS} )|\psi\rangle_{AB}|\varphi\rangle_{RS}\nonumber\\
&&\qquad =(\Delta_{RA} \otimes I_B)(A_A U_{RA}\otimes V_{BS})|\psi\rangle_{AB}|\varphi\rangle_{RS}\nonumber\\
&&\qquad =(\Delta_{RA} \otimes I_B)(A_A \otimes I_B)|\psi\rangle_{AB}|\varphi_0\rangle_{RS}\nonumber\\
&&\qquad =(R_R \otimes I_B)(A_A \otimes I_B)|\psi\rangle_{AB}|\varphi_0\rangle_{RS}~\nonumber\\
&&\qquad =(R_R \otimes I_B)(A_A U_{RA}\otimes V_{BS})|\psi\rangle_{AB}|\varphi\rangle_{RS}~,
\end{eqnarray}
and multiplying both sides by $I_A\otimes V_{BS}^{-1}$ gives
\begin{eqnarray}
\label{a_commute}
&& (U_{RA}A_A\otimes I_B )|\psi\rangle_{AB}|\varphi\rangle_{RS}\nonumber\\
 && =(R_R \otimes I_B)(A_A U_{RA}\otimes I_B)|\psi\rangle_{AB}|\varphi\rangle_{RS}~;
\end{eqnarray}
that is, acting on the state $|\psi\rangle_{AB}|\varphi\rangle_{RS}$, we may replace $U_{RA}A_A$ by $R_RA_A U_{RA}$. A similar argument shows that 
\begin{eqnarray}
\label{b_commute}
&& (I_A\otimes V_{BS}B_S)|\psi\rangle_{AB}|\varphi\rangle_{RS}\nonumber\\
 && =(I_A\otimes S_S)(I_A\otimes B_S V_{BS})|\psi\rangle_{AB}|\varphi\rangle_{RS}~,
\end{eqnarray}
where $S_S$ is a unitary transformation acting on Bob's ancilla.

Now we can use the commutation properties eq.~(\ref{a_commute}),(\ref{b_commute}) to determine how the superoperator ${\cal E}$ acts on $(A\otimes B)|\psi\rangle_{AB}$:
\begin{eqnarray}
&&(U_{RA}\otimes V_{BS})(A_A\otimes B_B)|\psi\rangle_{AB}|\varphi\rangle_{RS}\nonumber\\
&&\quad = (I_A\otimes V_{BS}B_S)(U_{RA}A_A\otimes I_B)|\psi\rangle_{AB}|\varphi\rangle_{RS}\nonumber\\
&&\quad = (R_R\otimes I_B)(I_A\otimes V_{BS}B_S)(A_AU_{RA}\otimes I_B)|\psi\rangle_{AB}|\varphi\rangle_{RS}\nonumber\\
&&\quad = (R_R\otimes I_B)(A_AU_{RA}\otimes I_B)(I_A\otimes V_{BS}B_S)|\psi\rangle_{AB}|\varphi\rangle_{RS}\nonumber\\
&&\quad = (R_R\otimes S_S)(A_AU_{RA}\otimes I_B)(I_A\otimes B_SV_{BS})|\psi\rangle_{AB}|\varphi\rangle_{RS}\nonumber\\
&&\quad = (R_R\otimes S_S)(A_A\otimes B_B)(U_{RA}\otimes V_{BS})|\psi\rangle_{AB}|\varphi\rangle_{RS}\nonumber\\
&&\quad = (R_R\otimes S_S)(A_A\otimes B_B)|\psi\rangle_{AB}|\varphi_0\rangle_{RS}\nonumber\\
&&\quad = \left[(A_A\otimes B_B)|\psi\rangle_{AB}\right]\left[(R_R\otimes S_S)|\varphi_0\rangle_{RS}\right]~.
\end{eqnarray}
We have shown that $(A_A\otimes B_B)|\psi\rangle_{AB}$ is an eigenstate of ${\cal E}$, which
completes the proof of Theorem 5.

\end{document}